\documentclass[fleqn,usenatbib]{mnras}

\usepackage[T1]{fontenc}

\DeclareRobustCommand{\VAN}[3]{#2}
\let\VANthebibliography\thebibliography
\def\thebibliography{\DeclareRobustCommand{\VAN}[3]{##3}\VANthebibliography}

\usepackage{graphicx}	
\usepackage{amsmath}	
\usepackage{amssymb}	
\usepackage{booktabs}   
\usepackage{multirow}

\usepackage{newtxtext,newtxmath}

\newcommand{\mini}{\mbox{$m_{\rm i}$}}
\newcommand{\feh}{\mbox{\rm [{\rm Fe}/{\rm H}]}}

\newcommand{\Msun}{\mbox{$\mathrm{M}_{\odot}$}}

\newcommand{\logg}{\mbox{$\log g$}}
\newcommand{\Teff}{\mbox{$T_{\rm eff}$}}
\newcommand{\gmag}{\mbox{${M}_{G}$}}
\newcommand{\gcolor}{\mbox{${G}_{\rm BP}-{G}_{\rm RP}$}}

\newcommand{\dmax}{\mbox{${d}_{\rm max}$}}
\newcommand{\pimin}{\mbox{${\pi}_{\rm min}$}}
\newcommand{\gbp}{\mbox{$G_{\rm BP}$}}
\newcommand{\grp}{\mbox{$G_{\rm RP}$}}

\newcommand{\astnoise}{\mbox{$N_\mathrm{ast}$}}

\newcommand{\fbin}{\mbox{$f_\mathrm{bin}$}}
\newcommand{\fresolved}{\mbox{$X$}}
\newcommand{\logtyr}{\mbox{$\log(t/\mathrm{yr})$}}

\title[Gaia HRD]{Dissecting the Gaia HR diagram within 200 pc}

\author[Dal Tio et al.]{
Piero Dal Tio$^{1,2}$,
Alessandro Mazzi$^{1}$\thanks{E-mail:  alessandro.mazzi.phd@gmail.com},
L\'eo Girardi$^{2}$,
Mauro Barbieri$^{3}$,
Simone Zaggia$^{2}$, \newauthor
Alessandro Bressan$^{4}$,
Yang Chen$^{5,1,6}$,
Guglielmo Costa$^{1,2}$,
Paola Marigo$^{1}$ 
\\
$^{1}$Dipartimento di Fisica e Astronomia Galileo Galilei, Universit\`a di Padova, Vicolo dell'Osservatorio 3, I-35122 Padova, Italy \\
$^{2}$Osservatorio Astronomico di Padova -- INAF, Vicolo dell'Osservatorio 5, I-35122 Padova, Italy \\
$^{3}$Universidad de Atacama, Instituto de Astronom\'{\i}a y Ciencias Planetarias, Copiap\'o, Chile \\
$^{4}$SISSA, via Bonomea 365, I-34136 Trieste, Italy\\
$^{5}$Anhui University, Hefei 230601, China\\
$^{6}$National Astronomical Observatories, Chinese Academy of Sciences, Beijing 100101, China
}

\date{Accepted XXX. Received YYY; in original form ZZZ}

\pubyear{2020}

\begin{document}
\label{firstpage}
\pagerange{\pageref{firstpage}--\pageref{lastpage}}
\maketitle

\begin{abstract}
We analyse the high-quality Hertzsprung--Russell diagram (HRD) derived from Gaia data release 2 for the Solar Neighbourhood. We start building an almost-complete sample within 200 pc and for $|b|>25^\circ$, so as to limit the impact of known errors and artefacts in the Gaia catalog. Particular effort is then put into improving the modelling of population of binaries, which produce two marked features in the HRD: the sequence of near-equal mass binaries along the lower main sequence, and the isolated group of hot subdwarfs. We describe a new tool, BinaPSE, to follow the evolution of interacting binaries in a way that improves the consistency with PARSEC evolutionary tracks for single stars. BinaPSE is implemented into the TRILEGAL code for the generation of ``partial models'' for both single and binary stellar populations, taking into account the presence of resolved and unresolved binaries. We then fit the Gaia HRD via MCMC methods that search for the star formation history (SFH) and initial binary fraction (by mass) that maximise the likelihood. The main results are (i) the binary fraction derived from the lower main sequence is close to 0.4, while twice larger values are favoured when the upper part of the HRD is fitted; (ii) present models predict the observed numbers of hot subdwarfs to within a factor of 2; (iii) irrespective of the prescription for the binaries, the star formation rate peaks at values $\sim\!1.5\times10^{-4}\Msun\mathrm{yr}^{-1}$ at ages slightly above 2~Gyr, and then decreases to $\sim\!0.8\times10^{-4}\Msun\mathrm{yr}^{-1}$ at very old ages.
\end{abstract}




\section{Introduction}
\label{sec:intro}

Fitting of color-magnitude diagrams (CMD) is nowadays the gold standard tool for deriving the star formation histories (SFH) of nearby galaxies. The basic idea in CMD-fitting is that the sub-pieces of galaxies are made by the addition of ``single-burst'' stellar populations (or partial models, PMs) of different masses, ages, and metallicities (and sometimes different extinctions), which in turn can be simply modelled from the basic theory of stellar structure and evolution, with the addition of simulated observational errors. Several methods in the literature share these same principles, although largely differing on the way the PMs are modelled and combined to identify a best-fitting model. Hundreds of galaxy regions within 1~Mpc have had their CMDs studied in this way \citep[see e.g.][]{dolphin02, tolstoy09, weisz11, gallart15, lewis15, rubele18}, allowing to identify main events in their history, and opening the way for the so-called ``near-field cosmology''.

Once reliable distances and extinctions allow us to convert apparent magnitudes into absolute ones -- hence allowing us to build the HR diagram (HRD) -- CMD-fitting methods can also be applied to stars in the Solar Neighbourhood. This has been done since the first release of Hipparcos catalogue \citep{hernandez00, bertelli01, vergely02, cignoni06}, although with several limitations: First, the photometric completeness of the Hipparcos input catalog was ensured only for very bright stars (roughly for $V\la7.3$). Second, samples built for the HRD analysis contained very few stars at the magnitude level of the oldest main sequence turn-offs, and they completely ignored the lowest main sequence made by unevolved stars, owing to the limited accuracy of Hipparcos parallaxes. An emblematic case is presented by \citet{hernandez00}, who limited their analysis to a volume-limited sample and hence could not derive the SFH for ages older than 3~Gyr. Although alternative methods exist to constrain the SFH in the Solar Neighbourhood \citep[such as the white dwarf luminosity function;][]{noh90, rowell13}, they are deemed to be more uncertain than the CMD-fitting involving stars in the main nuclear burning phases of stellar evolution.

The situation has dramatically changed with the release of Gaia data release 2 \citep[DR2;][]{gaiaDR2} which provided parallaxes more accurate than Hipparcos by a factor of $\sim\!20$. In addition, Gaia DR2 includes accurate and homogeneous photometry, with uncertainties of the order of millimags down to apparent magnitudes of $G\!\sim\!18$~mag. These improvements are evident in the beautiful HRDs illustrated in \citet{babu}.

In this paper, we aim at the interpretation of the Gaia DR2 HRD using the CMD-fitting method. Although some shortcomings in DR2 may still hamper a definitive analysis of the stellar content in the Solar Neighbourhood, its data clearly overcomes many limitations of the previous Hipparcos data and is of sufficient quality to allow us to verify the assumptions commonly used in the population synthesis models and CMD-fitting methods applied to external galaxies. A good overview of the possibilities opened by Gaia DR2 can be found in the independent CMD-fitting works by \citet{gallart19a,gallart19b}, \citet{mor19}, \citet{ruiz20} and  \citet{alzate20}. Also worth of mention are the attempts to improve the determinations of the stellar initial mass function (IMF) from \citet{sollima19} and \citet{hallakoun20}.

Specially important, in this regard, is the possibility of checking the prescriptions used to simulate unresolved binaries in CMD-fitting studies. Indeed, the Gaia DR2 HRD presents both a rich population of nearly equal-mass binaries distributed in a sequence parallel to the lower main sequence (which is commonly seen in HST data of Local Group dwarf galaxies and in star clusters, see \citealt{sollima07} for instance), and a sizeable population of hot subdwarfs (see \citealt{geier19}) originated from mass-transfer in close binaries \citep{han02,heber09}. Therefore, models of binary populations might aim at reproducing these HRD features too. Since the production rate of hot subdwarfs is expected to vary with the population age, reproducing their numbers cannot be separated from the problem of determining the best-fit SFH of a given volume-limited stellar sample. Conversely, the numbers of observed binaries might bring implications for the determination of the SFH, which are still to be fully explored in the literature. A preliminary investigation of the effect of unresolved binaries was recently reported by \citet{alzate20}, who find that ``ignoring the presence of unresolved binaries biases the inferred age-metallicity relation towards older ages and higher metallicities than the true values''.

In this paper, we aim to do an additional step in this direction, presenting a new formalism for the analysis of the Gaia HRD that is suited to calibrate parameters in binary population models, and at the same time allows us to estimate their impact on the SFH determinations. Subsequent papers will develop these methods further, then taking full advantage of the expected improvements in the data from EDR3 and DR3.

This paper is structured as follows. In Sect.~\ref{sec:data} we present our selection of Gaia DR2 data to build a -- as far as possible -- clean, volume-limited HRD for the Solar Neighbourhood. In Sect.~\ref{sec:models} we present the TRILEGAL population synthesis code used to model the Gaia HRD, concentrating on the new BinaPSE module to describe binary evolution and their products. More details are given in the Appendix~\ref{sec:examples}, with examples of binary evolution and a few simulations of simple stellar populations focusing on the binaries.  Sect.~\ref{sec:method} presents the modelling of the Gaia data in terms of a linear combination of simple stellar populations, and the method adopted to identify the best-fitting parameters. Sect.~\ref{sec:discuss} presents a few of the best-fitting models and the conclusions we can draw in terms of the recovery of SFH, and the presence of binaries. 

\section{Selecting Gaia DR2 data}
\label{sec:data}

\begin{figure*}
	\includegraphics[width=\textwidth]{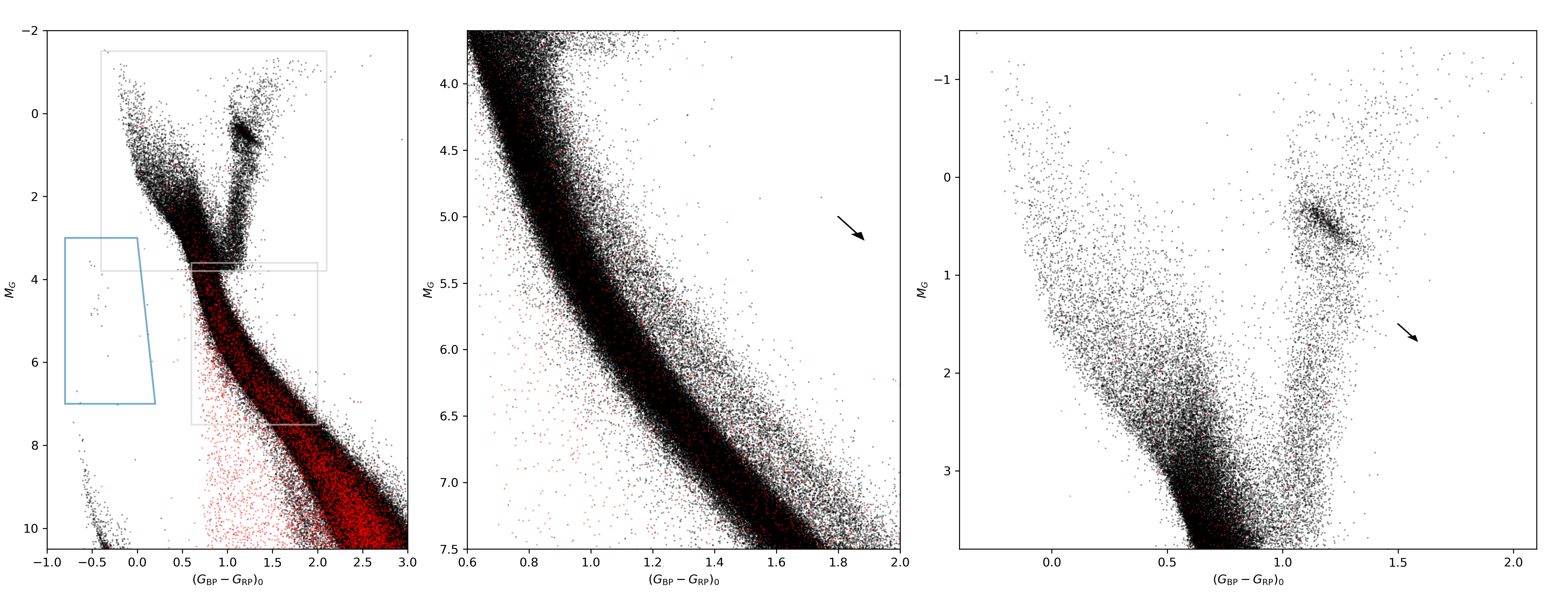}
    \caption{HRD for our initial sample. The left panel shows the \gmag\ versus $(\gbp-\grp)_0$ for all stars (black dots), overplotting in red the stars ultimately rejected because of their suspicious astrometry or photometry. The blue polygon delimits the region of hot subdwarfs. The central and right panels zoom at the lower main sequence and at the upper part of the HRD, respectively, better showing the   structures that will be used in our HRD-fitting work. On the latter panels, a reddening vector corresponding to $A_V=0.2$~mag is overplotted for comparison.}
    \label{fig:bighrd}
\end{figure*}

\subsection{Basic reasoning and initial catalog}
\label{sec:basic}

Our initial goal is to create a clean sample representative of the Solar Neighbourhood in the \gmag\ vs. \gcolor\ HRD from Gaia DR2. Most importantly, we aim at having something close to a ``complete volume-limited sample'', because it can be easily compared to the output of a population synthesis code -- which by definition generates all stars in a given volume. We list below the considerations that lead us to this sample. 

Under the assumption of small parallax errors, we can approximate the distance of a star given its parallax as $d=1/\pi$. We then discard stars farther than a maximum distance $\dmax$, which corresponds to a minimum parallax $\pimin = 1/\dmax$. The absolute magnitude \gmag\ and colour of any star in the sample are given by 
    \begin{eqnarray}
        \gmag &=& G + 5\log\pi + 5 - A_G \label{eq:absmag} \\ \nonumber
        (\gbp-\grp)_0 &=& \gbp-\grp - E(\gbp-\grp)
    \end{eqnarray}
where $G$ is the apparent magnitude and $A_G$ is the interstellar extinction. We recall that, in the limit of small extinction and for ``median-temperature stars'' such as the Sun, this extinction is related to the one in the $V$ band by $A_G=0.861\,A_V$. The colour excess is related to $A_V$ by $E(\gbp-\grp) = 0.421\,A_V$ (see \citealt{chen19}, assuming \citealt{cardelli89}'s extinction curve with $R_V=3.1$).

We then just need to define limits that ensure these relations are accurate and provide HRDs well populated of stars. Trial and error experiments have led us to the following choices:
\begin{enumerate}
    \item \dmax\ is set to 200~pc, or equivalently to a parallax threshold of $\pimin=5$~mas. With typical errors in DR2 parallaxes being 0.04~mas \citep{luri18}, this ensures distance errors typically smaller than $\sim\!1$~\%, and hence \gmag\ errors smaller than 0.02~mag. We note that the small offsets present in DR2 parallaxes, of the order of 0.03~mas \citep{lindegren, khan19,chan20}, become insignificant compared to this 5~mas threshold.
    \item Inspection of the 3D extinction maps from \citet{lallement18} reveals that for $d<200$~pc we have high extinction regions close to the Galactic Plane. We therefore limit the catalog to high galactic latitudes, with $|b|>25^\circ$. The remaining sample has $A_{G,\mathrm{median}}\simeq0.03$~mag, with 16\% and 84\% percentiles at $0.009$~mag and $0.06$~mag, respectively, and an absolute maximum value of $0.75$~mag. Importantly, less than 4 per cent of the stars in this ``nearby and out-of-plane'' sample have extinctions larger than 0.16~mag. Such low extinction values ensure that the corrections in eq.~\ref{eq:absmag} are accurate even considering the possible errors present in the extinction maps from \citet{lallement18}.
\end{enumerate}

A Gaia DR2 catalog with these $\pi$ and $b$ limits contains 1\,361\,767 stars. Fig.~\ref{fig:bighrd} presents its HRD between limits $-2<\gmag<10.5$ and $-1.0<(\gbp-\grp)_0<3.0$. As can be appreciated in the figure, it contains a significant number of stars in the main post-main sequence evolutionary phases. Just 3 star clusters from the \citet{kharchenko05} catalogue are contained within our limits, the most notable being Praesepe. 

\subsection{Culling the initial catalog}
\label{sec:culling}

\begin{figure*}
	\includegraphics[width=\textwidth]{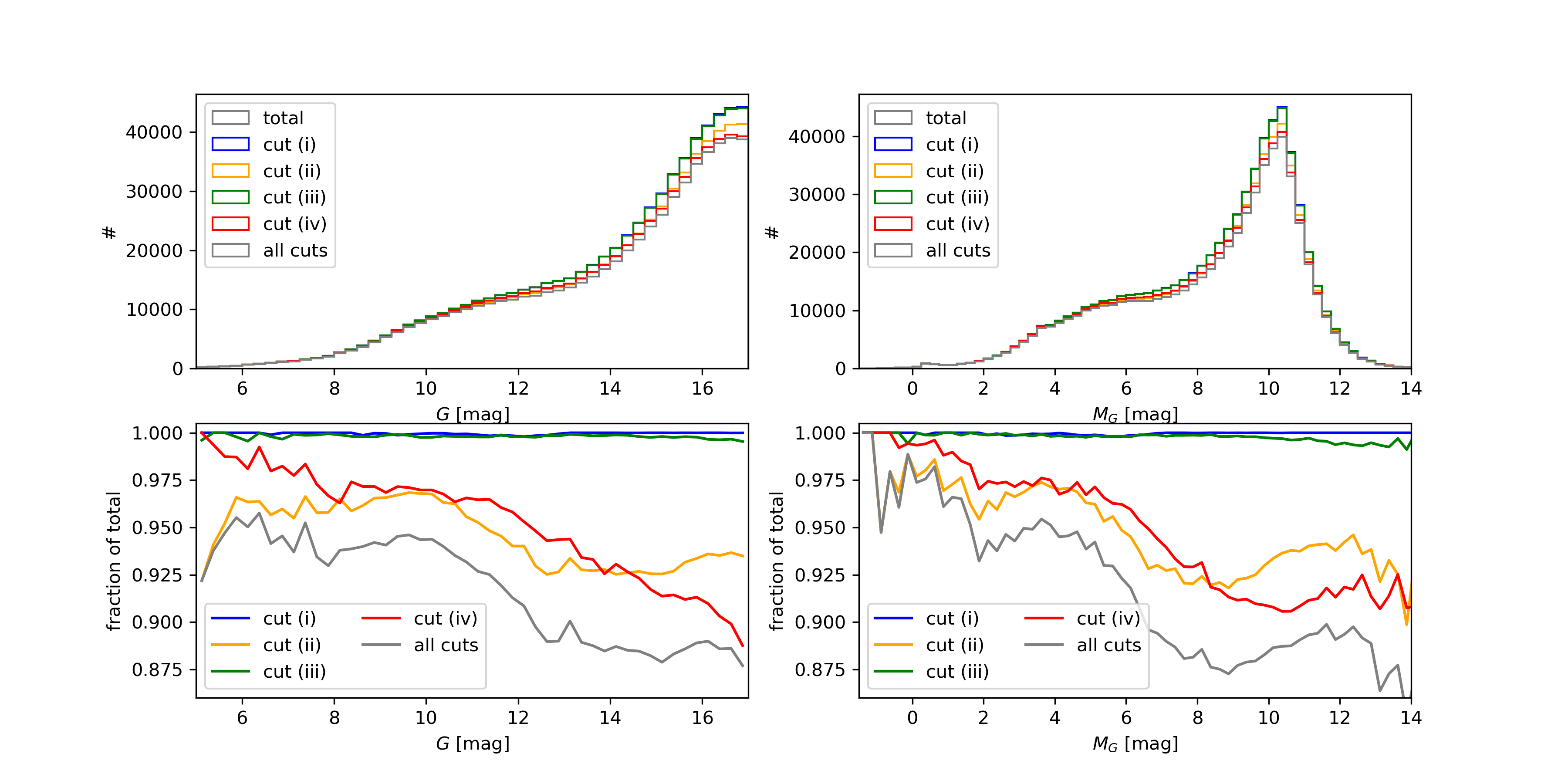}
    \caption{Top row: Number of stars in the initial catalog ($d<200$~pc and $|b|>25^\circ$), before and after the different cuts in astrometric and photometric quality described in Sect.~\ref{sec:culling}, as a function of both apparent (left panel) and absolute $G$ magnitude (right panel). Panels in the bottom row show the same in terms of fraction of stars remaining after the different cuts.}
    \label{fig:stars_lost}
\end{figure*}

The HRDs of Fig.~\ref{fig:bighrd} show all the beautiful features described by \citet{babu}, but also some artefacts caused by known problems in the DR2 astrometry and photometry. It happens that some stars falling in our sample have significant errors in their parallaxes and/or apparent magnitudes, so much that they appear in the wrong place of the HRD. They might even be spurious objects coming from outside our maximum distance. To deal with these objects, we define the following additional cuts:
\begin{enumerate}
\item \begin{verbatim} parallax_over_error > 5 \end{verbatim} \label{cut_parallax}
\item \begin{verbatim} astrometric_excess_noise < 1 \end{verbatim} \label{cut_astrometry}
\item \begin{verbatim} phot_bp_mean_flux_over_error > 10 
    AND phot_rp_mean_flux_over_error > 10\end{verbatim}  \label{cut_flux}
\item \begin{verbatim} phot_bp_rp_excess_factor < polyLine(bp_rp, 
    -0.56,1.307, 0.03,1.192, 1.51,1.295, 4.31,1.808) \end{verbatim} \label{cut_excess}
\end{enumerate}
Cuts \ref{cut_parallax} and \ref{cut_astrometry} aim to eliminate stars with unreliable parallaxes that passed the initial $\pi>5$~mas cut, while \ref{cut_flux} and \ref{cut_excess} eliminate stars whose photometry is either uncertain or suspiciously inconsistent between the $G$, \gbp, and \grp\ passbands. The last criterion is inspired by an example provided by \citet{taylor18}. 

Figure~\ref{fig:stars_lost} shows the fraction of stars retained after each one of these cuts, and after all the cuts, as a function of $G$ and $M_G$. As can be noticed, the most severe cuts are \ref{cut_astrometry} and \ref{cut_excess}, which cause the removal of a few percent of the stars over the interval $5\gtrsim G\gtrsim17$. The mean fraction of stars being retained after all cuts amounts to about 92\%, but approaches 94\% for the brighter stars, i.e. those with $G\la11$ and $M_G\la6$. 

Stars being eliminated by any of these cuts are plotted as red dots in the HRD of Fig.~\ref{fig:bighrd}. It can be seen that without these dots, we have a significantly cleaner HRD. In particular, we eliminate a large number of stars between the main sequence (MS) and white dwarf (WD) sequence corresponding to stars with bad astrometric solutions. The primary parameter describing these stars is a high  \verb$astrometric_excess_noise$, or $\astnoise$. While many of the stars eliminated may be real binaries for which the orbital motion explains the bad astrometric solution, stars with $\astnoise>1$ are concentrated along the Galactic Plane and clumped around sky locations for which the Gaia scanning is very poor so far, as discussed by \citet{lindegren}.
 
The drawback of applying these cuts is that we might be removing from the sample stars which in reality fulfil the $d<200$~pc and $|b|>25^\circ$ conditions. Conversely, there might be stars fulfilling these conditions that appear at very different apparent distances and hence did not even enter in our initial catalog. This situation might improve in the next Gaia data releases. 

Finally, let us consider the ranges of absolute magnitudes being comprised in the catalog.
\begin{enumerate}
    \item The faintest limit is set by the maximum distance and extinction in the sample. With $\dmax=200$~pc, the maximum true distance modulus is $\mu_0=6.505$~mag, and $>96$~\% of the stars have a maximum distance modulus in the $G$ band of $\mu<(6.505+0.16)=6.665$~mag. Therefore it is fair to say that a cut at $G<17$~mag represents a sample which is essentially complete for all absolute magnitudes $\gmag<(17-6.665)=10.335$~mag.
    \item At the brightest limit of $G=5$~mag, we either are dealing with the closest faint stars for which the extinction is null, or with the farthest bright stars, a small fraction of which might have some appreciable extinction. Anyway, if we consider the extinction correction as accurate, we have stars coming from absolute magnitudes equal to $\gmag=5-5\log d+5=10-5\log d$, that is, there is a one-to-one relation between the absolute magnitude of the observed stars, and the minimum distances being sampled.
\end{enumerate}
Summarising, the distances being sampled for the $5<G<17$ interval are:
\begin{equation}
    10^{-0.2 (M_G-10)}<(d/\mathrm{pc}) <200\,\,\,\, \mathrm{for~}\gmag<10.335 \mathrm{~mag} \,
\end{equation}
and the fraction of the total ``target volume'' is
\begin{equation}
    F = 1 - 10^{-0.6 (M_G-10)} / 200^3 \,.
\end{equation}
The latter is actually close to 1 for an ample interval of the $\gmag$ of interest. For instance, stars with $\gmag=3$~mag (close to the oldest main sequence turn-off) will be sampled at the entire $25<(d/\mathrm{pc})<200$ interval, which represents 99.8~\% of the target volume, while stars with $\gmag=0$ (in the red clump) will be at $100<(d/\mathrm{pc})<200$, which samples 87.5~\% of the target volume. These numbers suggest that we can even use $F$ as a completeness factor as a function of $\gmag$, as we do in the following. We note that $F$ falls to zero for stars brighter than $\gmag=-1.5$~mag. Moreover, we note that the accuracy of the extinction correction is relevant only for the samples observed prevalently at large distances, i.e., indicatively, for the stars with $-1.5<\gmag<0$.

Therefore, we can make use of the entire range of absolute magnitudes accessible with this catalog, from $-1.5<\gmag<10.365$. As can be seen in the HRD of Fig.~\ref{fig:bighrd}, this catalog provides a good sampling of the upper main sequence and of the entire RGB, RC included. At the faint end, it contains the brightest part of the beautiful WD cooling sequence delineated in \citet{babu}. It also samples very well the lower main sequence, down to initial masses of $\mini\lesssim0.45$~\Msun.

In order to ensure a completeness larger than 90\% and limit the impact of possible variations in the IMF of low-mass stars, we initially limit the sample to $M_G<7.5$~mag, as detailed below. 

\section{Models}
\label{sec:models} 

Now that our data has been set, our goal is to reproduce the Gaia DR2 HRD as a sum of single-burst stellar populations incorporating all the known errors and biases. To do that, we first need to build a set of single-burst stellar populations, hereafter called ``partial models'', or PMs. The subsequent step is to combine the PMs to provide a good fitting of the observed HRD. PMs can be produced in several forms, such as catalogues, luminosity functions, and Hess diagrams. In the following, we will refer to PMs mainly as Hess diagrams covering the limits defined from the Gaia DR2 HRD. However, the same PMs can be re-generated later in other forms, for the subsequent analyses.

\subsection{Single stars with TRILEGAL}

The primary tool we use to create PMs is TRILEGAL \citep{girardi05,girardi12}, a generic code to simulate stellar populations that has been widely used in the literature. It makes use of extensive libraries of stellar evolutionary tracks from PARSEC-COLIBRI teams (see Sect.~\ref{sec:binapse_code} below). Stars are sampled according to the initial mass function (IMF), and user-specified distributions of ages, metallicities, and distances. The simplest way of specifying these distributions is directly providing the SFH, intended as the star formation rate as a function of population age, SFR$(t)$, plus the age--metallicity relation, $\feh(t)$. Then, the simulated stars are converted into magnitudes via bolometric correction tables and extinction coefficients from the YBC code \citep[][see Sect.~\ref{sec:bc} below]{chen19}.

Until recently, TRILEGAL was able to simulate only single stars, and non-interacting binaries via the simple addition of the light output from the two binary components. In the following, we describe a new important addition to TRILEGAL, which allows us to introduce interacting binaries in the simulations. 

\subsection{BinaPSE: the new TRILEGAL module}
\label{sec:binapse_code}

The increasing relevance of interacting binary stars in modern astronomy motivated us to expand TRILEGAL capabilities by linking it with the BSE code \citep{hurley02}, a popular binary evolution code for population synthesis. However, it is not possible to use TRILEGAL and BSE by simply running them in sequence because they do not share the same evolutionary tracks, i.e. the predicted stellar loci on the HRD and stellar counts of single and binary stars would not be consistent one another. More precisely, TRILEGAL interpolates among pre-computed stellar evolutionary grids (i.e. it implements a grid-based method) to match the mass, the metallicity and the age of generated stars. On the other hand, BSE evolves the binary components by following analytic formulae which approximate a different set of evolutionary grids \citep[see][]{hurley2000}. Although the use of analytic formulae guarantees a very fast computation, we decided to revise BSE and to transform it into a grid-based code in order to satisfy our accuracy requirements and to make future changes of evolutionary grids much easier. The BSE revision led to the creation of a new TRILEGAL module that we named BinaPSE. BinaPSE shares with TRILEGAL the evolutionary grids and interpolation routines, but preserves the binary evolution methodology described in \cite{hurley02}. Therefore, when the two components interact via mass transfer or a common envelope (CE), at each time step, the remnants are determined as in the old BSE code, but they are located in the evolutionary grids as in the original TRILEGAL code. 

The evolutionary grids used in BinaPSE are:
\begin{itemize}
    \item the PARSEC v1.2S evolutionary tracks provided by \cite{bressan12}, revised as in \cite{bressan15} and extended as in \cite{chen15};
    \item new evolutionary tracks of naked helium stars, computed with the latest version of PARSEC \citep{costa19a,costa19b};
    \item the COLIBRI TP-AGB tracks \citep{Marigo_etal_13}, described by \cite{rosenfield16}, which determine the initial-final mass relation (IFMR); 
    \item up-to-date grids for post-asymptotic giant branch stars and carbon-oxygen white dwarfs (CO-WD) from the models described in \cite{Bertolami_2016} and \cite{Renedo_2010}, respectively.
\end{itemize}
No evolutionary grids have been included in TRILEGAL for helium white dwarfs (He-WD), for oxygen-neon white dwarfs (ONe-WD) and for neutron stars (NS). We plan to do this improvement in the next future, but in the meanwhile we use the same analytic formulae of \cite{hurley2000} to manage the evolution of these stars.

In Appendix~\ref{sec:examples} we provide examples of the evolution obtained with BinaPSE, compared with the one obtained with BSE. They show that the use of PARSEC tracks in the binary evolution changes not only the position of the main evolutionary features in the HR diagrams, compared to BSE, but also changes the final fate for a fraction of the binaries.

With TRILEGAL and BinaPSE we can perform many kinds of simulations, such as:
\begin{enumerate}
    \item Any Galaxy field, limited in apparent magnitude and/or in maximum distance from the Sun, with a given initial binary fraction.
    \item An object (galaxy or star cluster) at fixed distance with a given SFH and initial binary fraction. This is the option used to create the grid of PMs presented in Section~\ref{sec:PMgrid}.
    \item A set of binary stars with a given distribution of initial parameters. This option is used in Appendix~\ref{sec:examples}.
\end{enumerate}

\subsection{Probability distributions of initial parameters for binary systems}
\label{sec:probdist}

The mass of the single stars in TRILEGAL is simply derived from the IMF, which is assumed to be independent of all other parameters. In this work we adopt the IMF from \cite{kroupa02}. 

To simulate the binaries, we adopt three different prescriptions from the literature. The simplest one is also the most frequently used in the analyses of CMDs of nearby galaxies: It assumes that the binary components do not interact during their lifetimes, and that they present a flat distribution of mass ratios, $q=m_{\mathrm{i},2}/m_{\mathrm{i},1}$. In practice, in this case the binaries are made of two single stars selected from the same isochrone, and no orbital parameter is specified. Since the large majority of binaries with small mass ratios have secondaries too faint to compete with the light of the primary\footnote{The primary is defined as the initially more massive component.}, only binaries with a mass ratio above a given threshold, which we set at 0.7, are simulated.

More realistic distributions include the possibility of interacting binaries and hence prescriptions for other initial parameters, such as the period $P$ and the eccentricity $e$, across the entire range of possible mass ratios. As our reference distribution of this kind, we adopt the Monte Carlo model proposed by \citet{Eggleton2006}, which summarises decades of work on the statistics and evolution of binary systems: Let $m_{\mathrm{i},1}$ and $m_{\mathrm{i},2}\leq m_{\mathrm{i},1}$ be the initial masses of the two components of a binary system. The distributions of $P$, $q$, and $e$ are given by
\begin{equation}
    P=\frac{5\cdot 10^4}{m_{\mathrm{i},1}^2}\left(\frac{X_1}{1-X_1}\right)^\alpha,\qquad\alpha=\frac{3.5+0.13\,m_{\mathrm{i},1}^{1.5}}{1+0.1\,m_{\mathrm{i},1}^{1.5}} , \label{eq:period}
\end{equation}
\begin{equation}
    q=1-X^\beta_2,\qquad \beta=\frac{2.5+0.7\beta'}{1+\beta'}, \qquad\beta'=\frac{\sqrt{P}(m_{\mathrm{i},1}+0.5)}{10} , \label{eq:massratio}
\end{equation}
\begin{equation}
    e=X_3 ,  \label{eq:eccentricity}
\end{equation}
where $X_1$, $X_2$ and $X_3$ are independent random variables uniformly distributed in the $[0,\,1]$ interval. 

In addition, we implement the more recent distribution of binary masses and orbital parameters from \citet{moe17}. It combines empirical evidence from many different kinds of binaries and has a functional form far more complicated than the \citet{Eggleton2006} one -- for instance it includes a dependence of the binary fraction and mass ratio on the mass of the primary. We use the \citet{moe17} formulation to produce the binary populations associated with a \cite{kroupa02} IMF, in a way similar those produced in the \citet{Eggleton2006} case. We verify that this prescription produces a fraction of nearly-equal-mass binaries, or $F_\mathrm{twin}$ (defined as the initial fraction of binaries with $q>0.95$ among all binaries with $q>0.3$), of 0.08. This fraction is comparable with the empirical values between 0.03 and 0.1 derived by \citet{elbadry19} from main sequence wide binaries in Gaia DR2.

\subsection{Bolometric corrections}
\label{sec:bc}

Initially, TRILEGAL produces synthetic stars using only their main intrinsic properties, like the luminosity $L$, effective temperature \Teff, current mass $m$, surface gravity $g$, surface chemical composition, etc. Then, these parameters are used to compute the photometry in the sets of filters of interest, by applying the bolometric corrections and extinction coefficients extracted from the YBC database \citep{chen19}. The latter are interpolated inside a huge grid of pre-computed tables, derived from libraries of model atmospheres and their spectral energy distributions, which is largely based on model atmospheres from the ATLAS9 \citep{castelli03} and PHOENIX \citep{allard12} codes. Main parameters in the interpolation are \Teff, \logg, and other parameters related to the surface chemical composition, including the initial metallicity \feh, and the surface abundance of CNO elements for AGB stars. 

To deal with the products of close binary evolution, such a procedure has to be complemented with bolometric correction tables for He-rich stars. This is especially important for stars that lose their hydrogen-rich envelope through a CE phase, as illustrated in Section~\ref{sec:hestars} below.

To model the bolometric correction for the helium rich stars, we compute a grid of spectral energy distributions for pure hydrogen+helium atmospheres by using the Tlusty code \citep{Hubeny1988,Hubeny1995}. This new grid covers the ranges of $30\,000<\Teff/\mathrm{K}<100\,000$, $5<=\logg [\mathrm{cm\,s^{-2}}]\leq9$, and $X/Y=[0.,0.1,...,1.0]$. Bolometric corrections and extinction coefficients for all filter sets of interest are then produced with the YBC code \citep{chen19}. Whenever BinaPSE produces a He-rich hot star, these tables are interpolated using $(\log\Teff,\logg, X/Y)$ as the interpolation parameters.

\section{Methods}
\label{sec:method} 

\begin{table}
    \caption{Some properties of the partial models. The first 3 columns refer to all PMs used in this work, while the other columns refer to the PMs along the reference AMR defined in Sect.~\ref{sec:ramr}. $N_\mathrm{HSds}$ refers to the number of hot subdwarfs defined as in Fig.~\ref{fig:bighrd}, derived with a constant SFR$(t)=1\Msun\mathrm{yr}^{-1}$, from the BinaPSE code and for the \citet{Eggleton2006} distribution of initial binary parameters. For comparison, $N_\mathrm{HSds}^\mathrm{BSE}$ presents results from the BSE code. }
    \centering
    \makebox[\columnwidth][c]{%
    \begin{tabular}{l|lllllll}
        \hline
        $i$ & \logtyr\ & $\Delta t$ & $Z_0$ & $N_\mathrm{HSds}$ & $N_\mathrm{HSds}/\Delta t$ & $N_\mathrm{HSds}^\mathrm{BSE}$ \\ 
            & interval &   (yr)     &   (initial)    &                   \\
        \hline
         1 & 6.6--7.1  & 8.61e+06 & 0.02146 &     0.0 & 0.0      &     0.0 \\ 
         2 & 7.1--7.3  & 7.36e+06 & 0.02145 &    27.0 & 3.66e-06 &     0.0 \\ 
         3 & 7.3--7.5  & 1.17e+07 & 0.02144 &   170.9 & 1.46e-05 &     0.0 \\ 
         4 & 7.5--7.7  & 1.85e+07 & 0.02142 &   180.6 & 9.76e-06 &    45.2 \\ 
         5 & 7.7--7.9  & 2.93e+07 & 0.02139 &   823.0 & 2.81e-05 &     0.0 \\ 
         6 & 7.9--8.1  & 4.65e+07 & 0.02134 &  1814.3 & 3.91e-05 &    56.6 \\ 
         7 & 8.1--8.3  & 7.36e+07 & 0.02126 &  9161.2 & 1.24e-04 &  1167.8 \\ 
         8 & 8.3--8.5  & 1.17e+08 & 0.02114 & 20369.2 & 1.75e-04 &  4130.7 \\ 
         9 & 8.5--8.7  & 1.85e+08 & 0.02096 & 42909.9 & 2.32e-04 &  8356.4 \\ 
        10 & 8.7--8.9  & 2.93e+08 & 0.02066 & 81260.1 & 2.77e-04 & 21120.5 \\ 
        11 & 8.9--9.1  & 4.65e+08 & 0.02020 & 39147.3 & 8.73e-05 & 27233.0 \\ 
        12 & 9.1--9.3  & 7.36e+08 & 0.01949 & 13487.1 & 1.83e-05 & 17983.6 \\ 
        13 & 9.3--9.5  & 1.17e+09 & 0.01842 & 17098.8 & 1.47e-05 &  4274.6 \\ 
        14 & 9.5--9.7  & 1.85e+09 & 0.01684 &  2259.1 & 1.22e-06 &  6776.3 \\ 
        15 & 9.7--9.9  & 2.93e+09 & 0.01461 &  3578.9 & 1.22e-06 &     0.0 \\ 
        16 & 9.9--10.1 & 4.65e+09 & 0.01167 &  5674.0 & 1.22e-06 &  5673.9 \\ 
        \hline
        $\sum_i$ & & 1.26e+10 & & 237961.4 & & 96818.4 \\
        \hline
    \end{tabular}}
    \label{tab:PMs}
\end{table}

\subsection{The grid of partial models}
\label{sec:PMgrid}

Although any age and metallicity values can be simulated with our codes, it is convenient to define PMs that cover a limited interval of such ages and metallicities. For the present work, we define PMs in 16 age intervals comprising all $\logtyr$ values between 6.9 and 10.1. Their properties are listed in Table~\ref{tab:PMs}. Each PM covers age intervals of $\Delta\log t=0.2$~dex, with the exception of the first one which covers 0.5~dex. PMs are distributed along a given reference age-metallicity relation (RAMR). Our initial guess, to be refined in Sect.~\ref{sec:ramr} below, is to assume an almost-flat RAMR, where models have a metallicity close to solar and a constant Gaussian metallicity dispersion of $\sigma=0.1$~dex, at all ages. This guess is motivated by the comparison with stellar models in Fig.~\ref{fig:rc}, and also by the correlation between ages and mean metallicities derived from nearby samples of red giants observed spectroscopically \citep[e.g.][]{haywood13,feuillet16,lin20}. Only for very old stars there are hints of the mean metallicity shifting to values a few tenths of dex smaller than the solar value. 

PMs are built separately for single and binary populations, for the same age-metallicity bins, and for the same total initial mass of stars. This approach allows us to define the fraction of stellar mass that is used to form binaries, \fbin. This quantity is independent of age. A complex population containing binaries can be derived by simply adding single and binary PMs with
\begin{equation}
    \mathrm{PM}_i = (1-\fbin)\,\mathrm{PM}_{\mathrm{sin},i} + 
    \fbin\, \mathrm{PM}_{\mathrm{bin},i}
    \label{eq:PMsinbin}
\end{equation}

In this work, binaries are initially considered as unresolved binaries, and their magnitudes represent either the light of both components added together, or the light from already-merged stars. This assumption will be improved starting from Sect.~\ref{sec:unresolved} below.

In addition, we compute a PM corresponding to the halo stars in the Solar Neighbourhood. This component is simply derived by using the standard calibration of the halo in TRILEGAL \citep{girardi12}, inside the 200-pc distance limit and for $|b|>25^\circ$. This is a component that can be kept fixed in our method, since it is essentially an isotropic component whose mean density has been well calibrated using faint star counts in deep surveys \citep[e.g.][]{groenewegen02}. Its age and metallicity distribution is also sufficiently well known and it does not need a recalibration. Moreover, it makes a so small contribution to the local star counts (about 4000 stars inside our magnitude limits), that possible errors in its description are not critical. 

\subsection{Including photometric and astrometric errors}

We do our best to incorporate the known errors in Gaia DR2 in our models. We start by deriving the median values of the errors in all observables involved in our work (namely $G$, \gbp, \grp, and $\pi$), as a function of the apparent magnitude $G$. They are computed using the quantities \verb$phot_[F]_mean_flux_error$  and \verb$parallax_over_error$ in the Gaia DR2 catalogue, where \verb$[F]$ stands for the three different filters.

Then, we generate synthetic samples of stars uniformly distributed within in a sphere of radius 200~pc, and uniformly distributed across the Gaia HRD (within limits $-2<\gmag<12$ and $-1<\gcolor<4$). For each fake star, 1$\sigma$ errors are selected from the observed relations involving the apparent magnitude, and then used to generate errors from Gaussian distributions. Errors in apparent magnitudes and parallaxes are then converted into $M_G$ and \gcolor\ errors. This process is repeated millions of times to give us a distribution of errors to be applied to the original PMs, at every small cell of the Hess diagram. This process is akin to the ``artificial star tests'' usually performed in the CMD-fitting of external galaxies and star clusters -- but with the significant difference that we use the likely errors as derived from the Gaia DR2 catalog, instead of inserting the fake stars in the original Gaia images. 

The comparison with the original error-free PMs reveals that the impact of simulated errors is quite modest. In the entire $M_G$ range, the only stars to be significantly spread in the Hess diagram are the brightest RGB and TP-AGB stars, which are nearly absent in the $d<200$~pc sample, therefore having negligible weight in the HRD-fitting process to be discussed below.

We then repeat the generation of synthetic samples, now extending the distances out to 250~pc. This simulation allows us to estimate the amount of stars that, being actually out of the 200~pc distance limit, turn out to be ``scattered'' inside this limit due to the errors in parallax -- and vice-versa. Since the scattering from outside-in is more frequent than the one from inside-out, this effect could lead to an overestimation of the stellar density in the 200~pc volume, especially at fainter magnitudes for which the parallax errors are large. However, we verified that this effect is less of a problem for us: in the entire magnitude interval $-1.5<\gmag<10$, just 0.6 per cent of stars are likely to be misplaced in this way. Moreover, this effect is almost symmetrical: the numbers of stars moving inwards is just 2.5 per cent larger than the number of stars moving outwards. Only for very faint magnitudes (say for $\gmag>15$, which is far fainter than the limits adopted in our analyses) the effect becomes relevant, owing to the increased parallax errors. 

\subsection{Defining a model and its likelihood}

Summarising, a total of $i=1,\ldots,16$ age bins are considered. Every one of these PMs was rescaled so as to correspond to the initial mass produced in that age interval by a constant star formation rate of 1~$\Msun\, \mathrm{yr}^{-1}$. Moreover, single and binary PMs are combined with Eq.~\ref{eq:PMsinbin}.  Under these conditions, a model can be defined as
\begin{equation}
    \mathrm{M} = \mathrm{PM}_0 + \sum_i a_i \,\mathrm{PM}_i
    \label{eq:PMstot}
\end{equation}
where the coefficients $a_i$ give the star formation rate as a function of age, directly in units of $\Msun\, \mathrm{yr}^{-1}$. $\mathrm{PM}_0$ is the halo PM, which is kept fixed. The fact that this is a simple linear combination makes the computation extremely fast, even for long Markov chains (see below).

As a variation to this scheme, we can use sets of PMs computed, at all ages, with small shifts in metallicity, $\Delta\feh$, with respect to the RAMR. Usually, we adopt $\Delta\feh=0.12$~dex, and compute PMs for three multiple values of $\Delta\feh$ above the RAMR, and for three below the RAMR -- hence spanning a total range of 0.72~dex in metallicity at every age. These sets of PMs are tagged as $\mathrm{PM}_i^{+1}$, $\mathrm{PM}_i^{+2}$, $\mathrm{PM}_i^{+3}$, $\mathrm{PM}_i^{-1}$, $\mathrm{PM}_i^{-2}$ and $\mathrm{PM}_i^{-3}$. To use them, Eq.~\ref{eq:PMstot} is modified to
\begin{equation}
    \mathrm{M} = \mathrm{PM}_0 + \sum_i a_i 
        \left[ (1-f_i)\, \mathrm{PM}_i^- + f_i\, \mathrm{PM}_i^+ \right]
    \label{eq:PMstotmet}
\end{equation}
where $\mathrm{PM}_i^-$ and $\mathrm{PM}_i^+$ are the two PMs whose metallicities, $\feh^+$ and $\feh^{-}$, bracket the desired one at that age, and $f_i=(\feh-\feh^{-})/(\feh^+-\feh^{-})$. In this way, we simulate small changes in metallicity by means of linear combinations of PMs, keeping the computational speed in the calculation of $\mathrm{M}$. We define 16 coefficients $z_i$, aimed describe the changes in metallicity at every age bin.

That said, our codes produce Hess diagrams with the same limits as the observed one. For the data-model comparison, we adopt the following definition of likelihood ratio derived from a Poisson distribution:
\begin{equation}
    \ln \mathcal{L} = \sum_k \left( O_k-M_k - 
        O_k \ln\frac{O_k}{M_k} 
    \right)
    \label{eq:likelihood}
\end{equation}
where $O_k$ and $M_k$ are the observed and model star counts, respectively, in the HRD bins of index $k$ \citep{vanhollebeke09, dolphin02}. For all HRD bins in which there is a significant number of observed and model stars, results are similar to half of the classical $\chi^2$ (or Gaussian likelihood ratio) where the standard deviation is given by the square root of the observed star counts \citep[see the discussion in][]{dolphin02}. 

\subsection{Finding the best-fit model}

Equation~\ref{eq:PMstotmet} comprises a maximum of 33 coefficients to be determined (16 $a_i$, 16 $z_i$, and \fbin). Actually, this number can be reduced by imposing that the same $z_i$ is valid for many age bins -- since, for instance, young populations are expected to be chemically homogeneous. We initially simplify the problem imposing that \textit{all} age bins have the same $z_i$, hence reducing the number of coefficients to 18.
The problem is then easily solvable by means of a Nelder-Mead minimization of $-\ln \mathcal{L}$. We use the routine taken from \citet{nr}, which typically converges to the likelihood maximum in a question of seconds.

There is no warranty that the solution found by the Nelder-Mead step is not trapped into a local minimum. To find more likely solutions and estimate the errors, we proceed with a Markov Chain Monte Carlo method. Two new codes were used in this case: either the \verb$trifit$ quick code in C, which implements a Metropolis-Hastings \citep{metropolis53} algorithm  following the guidelines by \citet{hogg18}, or the \verb$trimcmc$ python code built around the \verb$emcee$ package by \citet{emcee}, which appears to more efficiently explore the space of parameters in the case of long Markov chains. In the first case we start with $\sim\!500$ walkers from the solution indicated by the Nelder-Mead step, while in the second case we pick up from the solution determined by \verb$trifit$ using $100$ walkers. 
As a rule, more than 2000 steps are followed in the explorative runs made with \verb$trifit$, while 50000 steps are performed in the final runs with \verb$trimcmc$ presented below. The final models generally represent a visibly better fit to the observations than those resulting from the Nelder-Mead step. We use the final position of the walkers to derive the median, 68 and 95 per cent confidence intervals for all model parameters. We also derive a ``best-fit Hess diagram'', which is the simple average of the Hess diagrams obtained from the final positions of the walkers. This latter is used to illustrate the final solutions in the following.

Two additional features are turned on during the MCMC step: The first are age-dependent variations in $z_i$, which are aimed to explore possible changes in the adopted RAMR. After a few experiments with different prescriptions, we opt for a simple scheme in which there are just 3 $z_i$ coefficients to be varied, representing metallicity changes at the two age extremes (0 and 12.6 Gyr) and at an intermediate age of 2 Gyr. For any other age value, metallicity is linearly interpolated among these values. In this way, we adopt smooth changes in metallicity occurring over scales of Gyr, avoiding rapid changes that could be considered as unrealistic. The second change is to allow models $\mathrm{M}$ to be displaced in the HRD by quantities $\Delta\gmag$ and $\Delta(\gbp-\grp)_0$. These displacements are intended to simulate systematic offsets in the zeropoints in the Gaia photometry -- offsets that can be present either in the data, or in the models if the bolometric corrections are calculated with the wrong filter transmission curves. Given the succession of different filter transmission curves and corrections provided for Gaia DR2 \citep[][]{evans18, weiler18, maiz18}\footnote{\url{https://www.cosmos.esa.int/web/gaia/dr2-known-issues}}, and now for Gaia EDR3\footnote{\url{https://www.cosmos.esa.int/web/gaia/edr3-passbands}}, we cannot exclude that such offsets are present in our analysis. Fortunately, we find that the derived $\Delta\gmag$ and $\Delta(\gbp-\grp)_0$ never exceed 1 bin size of our Hess diagrams, which means they are always less than a few hundredths of magnitude. 

In the following, we will refer to the final distribution of $a_i$ and $z_i$ coefficients as our SFH solution. This solution can be easily decomposed into the star formation as a function of age, SFR$(t)$, and in the age-metallicity relation (AMR), $\feh(t)$, and their confidence intervals. 

\subsection{Metallicity range and definition of the RAMR}
\label{sec:ramr}

\begin{figure}
	\includegraphics[width=\columnwidth]{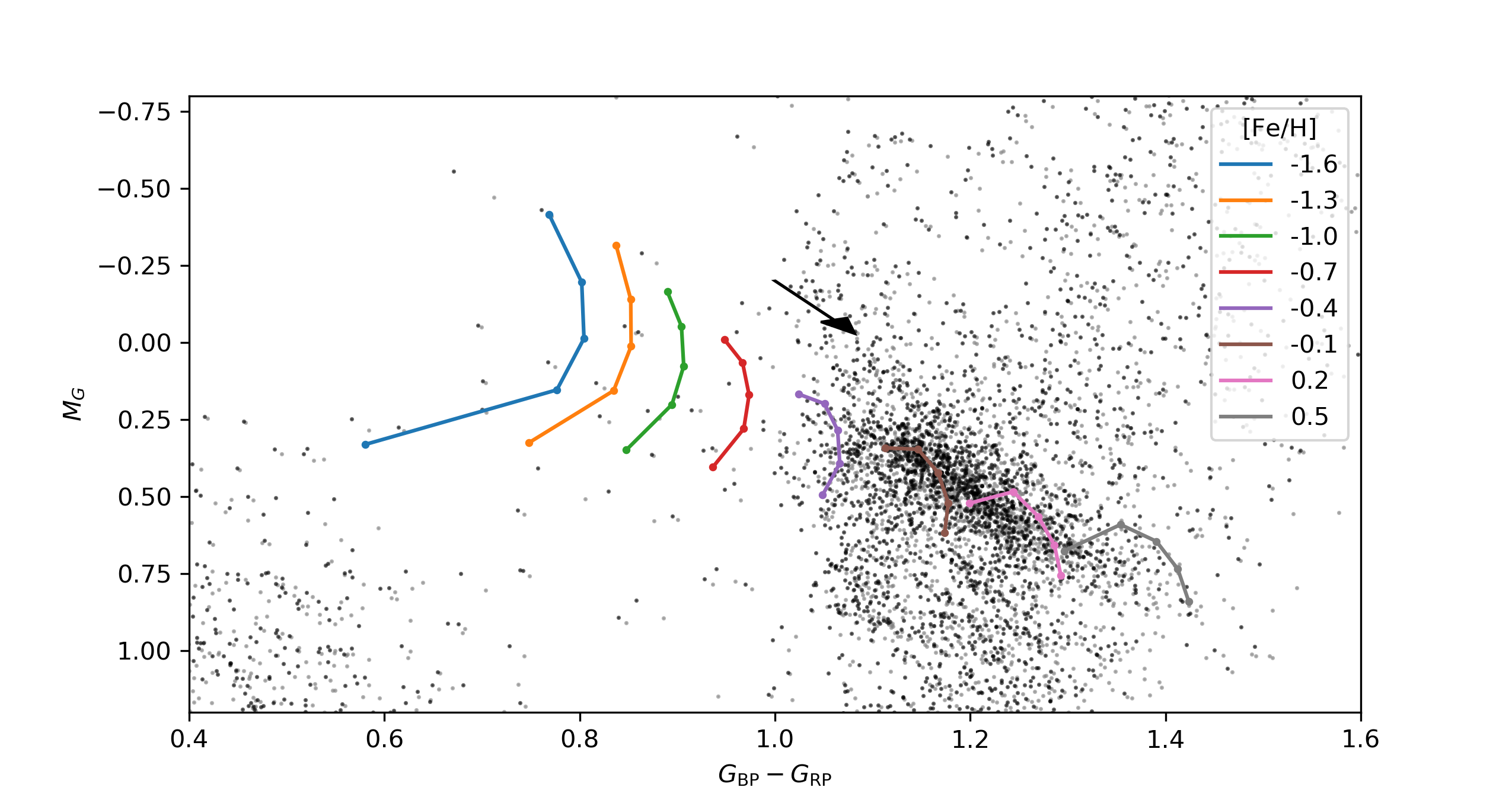}
    \caption{Zoom in the RC region of the HRD. The extinction-corrected data is marked with black dots, the original data with gray dots.  For comparison, the arrow illustrates the reddening vector corresponding to a range in extinction of $\Delta A_V=0.2$~mag, which exceeds the correction applied to the bulk of the data. The coloured lines present the mean location of the RC for several values of age and metallicity, as derived from PARSEC isochrones: Each line is for a different metallicity (as in the legend), and comprises 5 values of $\log\mathrm{(age/yr)}$ going from 9.3 to 10.1 at steps of 0.2 dex (the faintest point in a sequence is the oldest one). It can be seen that the RC stars follow a slope roughly consistent with a range of metallicities, with the bulk of RC stars having colours compatible with the $-0.4<\feh<+0.5$ interval.}
    \label{fig:rc}
\end{figure}

Figure~\ref{fig:bighrd} presents the Gaia data after the correction by extinction using the \citet{lallement18} 3D extinction map. The RC appears about 0.3 mag wide in color, and with a slope very similar to the reddening vector. This is illustrated with more detail in the zoomed HRD of Fig.~\ref{fig:rc}. 
As can be appreciated, the reddening correction has a marginal effect in this HRD. In the latter figure, we overplot stellar models giving the mean position of the RC for several metallicities and in sequences covering very wide age intervals. This comparison clarifies that the RC slope closely follows the mean slope expected for stellar populations spanning a range of metallicities, which then appears as the main factor driving the RC spread. Models in the range $-0.4<\feh<0.5$ suffice to explain the bulk of RC stars. We take this a first indication of the metallicity range needed to model these data. 

\begin{figure}
    \includegraphics[width=\columnwidth]{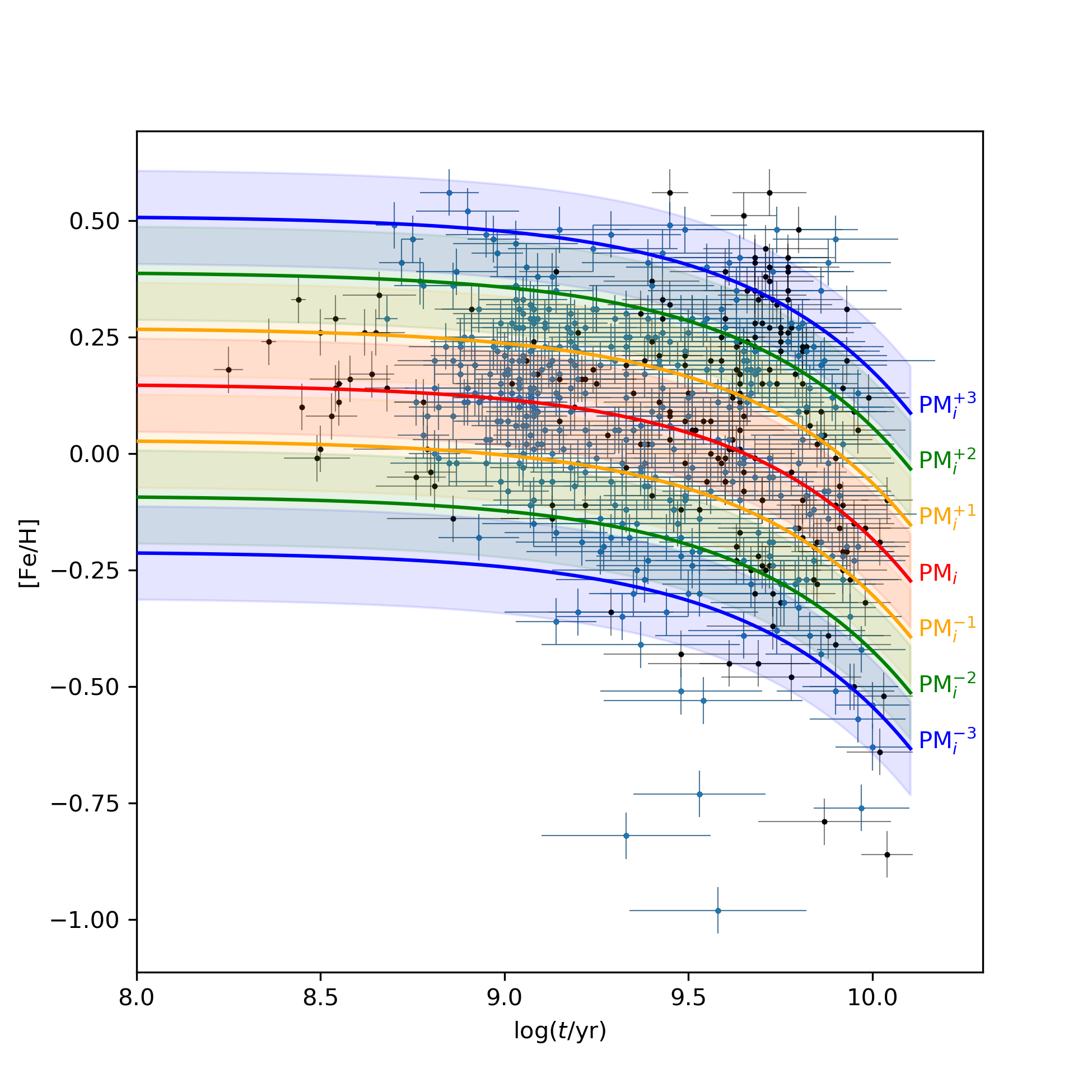}
    \caption{The ages and metallicities adopted in this work, compared to empirical data.
    The central red line shows the mean RAMR from eq.~\ref{eq:ramr}, together with the $\pm1\sigma$ interval of its Gaussian width (light red-shaded area). Similar sequences are then drawn for the PM sequences labelled as PM$_i^{\pm1}$, PM$_i^{\pm2}$, and PM$_i^{\pm3}$ (in orange, green and blue, respectively). We note that our seven PM sequences completely fill the area between the PM$_i^{\pm3}$ limits, with adjacent PMs largely overlapping in metallicity. Also, all sequences extend to younger ages, not shown in this figure. The dots with error bars represent the 598 giants and subgiants with ages and metallicities measured by \citet{feuillet16} that are inside our 200~pc distance limit (according to Gaia DR2 parallaxes). The blue dots identify the 384 stars with $0.8<\gcolor<1.6$ and $1.2<\gmag<-0.4$, that is, around the red clump region of the HRD (see Fig.~\ref{fig:rc}). 97 per cent of these stars are covered by the $\pm1\sigma$ limits of our PMs.}
    \label{fig:ramr}
\end{figure}

We then explore models with a large range of different RAMRs, centred on this metallicity range, verifying whether they converge to acceptable solutions. This is performed using the entire $\gmag<7.5$~mag interval, and using the code \verb$trifit$ limited to 2000 steps. Among many different possibilities tested, we soon opted for a family of models and solutions in which the reference AMR linearly varies as a function of age, with a slope $\alpha$ and the metallicity being solar at the age of the Sun's birth\footnote{We note that this is an approximation, because the Sun was not born with its present surface composition. According to the PARSEC tracks the present Sun with $(Z_\odot,Y_\odot)=(0.0152,0.2485)$ derives from a star with initial $(Z,Y)=(0.01774,0.28)$ \citep[cf.][]{bressan12}, and hence with $\feh\simeq0.02$~dex. This small offset can be taken into account in our method, by means of the metallicity shifts that are derived at all ages.}:
\begin{equation}
    \feh = \alpha(t-4.5\mathrm{Gyr})
    \label{eq:ramr}
\end{equation}
Not surprisingly, we find that models in which there is no metallicity change, and in which the RAMR total metallicity changes by more than 0.6~dex between young and old ages, simply do not provide good fits to the data. In contrast, models with $\alpha$ values comprised between $-0.2$ and $-0.6$ dex/12~Gyr present a clear marked decrease in their $-\ln\mathcal{L}$, achieving a final AMR in which the old disk is $\sim\!0.3$~dex more metal poor than the present young disk. 
We therefore define our RAMR to be the one with $\alpha=-0.4$~dex/12Gyr. We emphasize that the adopted PMs cover a $\pm0.36$~dex wide interval around this RAMR, therefore this choice by no means represent a strong limitation to the fitting solutions to be discussed below.

The RAMR, and the total range of metallicities explored around it, are plotted in Fig.~\ref{fig:ramr}. For comparison, we overplot the stellar sample with APOGEE spectroscopic metallicities\footnote{We use the [M/H] from APOGEE, which closely corresponds to [Fe/H].} and their Bayesian-derived ages from \citet{feuillet16}, limited to stars within 200~pc. We verify that these data uniformly sample the $\gcolor>1$ interval of RC stars in the Solar Neighbourhood, which was illustrated in Fig.~\ref{fig:rc}. One can appreciate that our PMs almost entirely cover the range of ages and metallicities indicated by APOGEE. These metallicities are concentrated at values around $+0.2$~dex, with just 7 out of 384 RC stars falling in the $-1<[\mathrm{M/H}]<-0.7$ metal-poor tail of the distribution. We also recall that the metallicities from \citet{feuillet16} are confirmed by the most recent APOGEE data releases \citep[see figure 3 in][]{feuillet19}, and that the scarcity of giants in the Solar Neighbourhood with metallicities smaller than $-0.7$~dex is confirmed by recent GALAH data \citep[see figure 14 in][]{nandakumar20}.

An independent confimation of the limited range of metallicities that needs to be explored, is given by the metallicity distribution of long-lived G and K dwarfs in the immediate Solar Neighbourhood. Different versions of these distributions \citep{rochapinto96, haywood01, haywood13, kotoneva02, casagrande11} indicate that the bulk of \feh\ values is comprised between $-0.5$ and $+0.5$~dex, with just a tiny fraction of dwarfs extending down to $\feh=-1$~dex.

To better quantify the fraction of metal-poor stars that might be present in our Gaia sample, we can reason as follows: There are at most 48 stars observed in the box with $1>\gmag>-0.7$, $0.6<\gcolor<1$, which is the HRD region corresponding to RC stars with metallicities between $-0.7$ and $-1.6$ dex. This number is to be compared to the 2009 redder RC stars, i.e.\ with $1<\gcolor<1.4$, which have higher metallicities. The RC corresponds to a major evolutionary stage (the core helium burning) appearing in all stellar populations older than $\sim1.5$~Gyr at a rate of {\em at least} 0.5 RC stars per every $10^3$~\Msun\ of star formation \citep[see figure 5 in][]{girardi16}. This implies that the above-mentioned 48 stars -- if they are really old RC stars, and not younger RC stars, or even younger stars crossing the Hertzsprung gap -- set an upper limit of $3\times10^4$~\Msun\ to the mass of stellar populations formed with $\feh<-0.7$~dex and presently in the Solar Neighbourhood. As we will see later in this section, this is a small fraction of the $\sim\!10^6$~\Msun\ total mass formed that can be estimated from the HRD fitting of our Gaia sample. Moreover, a $3500$~\Msun\ of metal poor populations are anyway included in our PM$_0$ model for the halo stars (Sect.~\ref{sec:PMgrid}).

Therefore, we can assume that our set of PMs cover the properties of the bulk of Solar Neighbourhood stars, just missing the trace populations with metallicities smaller than $-0.7$~dex -- which are, at least partially, included in the PM$_0$ model.


\begin{table*}
	\caption{Results of HRD fitting using different input models, parameters, and HRD limits}
	\label{tab:chi2}
	\begin{tabular}{llcccccccc} 
		\hline
		 model &
		 HRD             & binary           & binary           & $-\ln\mathcal{L}$ & \fbin & $\Delta(\gbp\!-\!\grp)_0$ & $\Delta\gmag$ & N of hot & comments\\
		 family & region$^1$      & evolution        & parameters$^2$        &              &      &  (mag) & (mag)  &  subdwarfs & \\
		\hline
		\multirow{4}*{\textbf{1}}
		& D      & BinaPSE & E06, unresolved  &  876 & 0.411 [0.392,0.426] & -0.0005 & -0.049 &  8.6 & only upper MS \\ 
		& C      & BinaPSE & E06, unresolved  & 1897 & 0.635 [0.620,0.654] & -0.0003 & -0.047 & 13.5 & upper MS+giants \\ 
		& A      & BinaPSE & E06, unresolved  & 5339 & 0.423 [0.420,0.428] & -0.009  & -0.024 &  9.1 & all HRD \\ 
		& B      & BinaPSE & E06, unresolved  & 1934 & 0.228 [0.219,0.235] & -0.032  & -0.080 &  1.4 & only lower MS \\ 
		\hline
		\multirow{4}*{\textbf{2}}
		& D      & BinaPSE & E06, resolved  & 889 & 0.898 [0.846,0.952]] & 0.009 & -0.086 & 19 & \\ %
		& C      & BinaPSE & E06, resolved & 2246 & 0.993 [0.981,0.998] & -0.003 & -0.071 & 22 & \\ 
		& A      & BinaPSE & E06, resolved  & 6816 & 0.9989 [0.9981,0.9994] & 0.021 & 0.023 & 26 & \\ %
		& B      & BinaPSE & E06, resolved  & 2143 & 0.408 [0.385,0.605] & -0.035 & -0.076 & 1.9 & \\ 
		\hline
		\multirow{3}*{\textbf{2a}}
		& D      & BinaPSE & E06, resolved  & 926 & 0.408 & 0.009 & -0.083 & 8.3 & fixed $\fbin$ \\ %
		& C      & BinaPSE & E06, resolved & 2368 & 0.408 & -0.002 & -0.068 & 8.3 & fixed $\fbin$ \\ %
		& A      & BinaPSE & E06, resolved  & 9296 & 0.408 & 0.014 & -0.050 & 7.9 & fixed $\fbin$ \\ %
		\hline
		\multirow{4}*{\textbf{3}}
		& D      & BinaPSE & MDS17, resolved  &  889 & 0.789 [0.733,0.858]] & 0.009 & -0.087 & 20 & \\ 
		& C      & BinaPSE & MDS17, resolved  & 2334 & 0.860 [0.834,0.881] & -0.002 & -0.062 & 22 & \\ 
		& A      & BinaPSE & MDS17, resolved  & 11587 & 0.973 [0.969,0.976] & 0.006 & 0.032 & 24 & \\ 
		& B      & BinaPSE & MDS17, resolved  & 2651 & 0.370 [0.354,0.389] & -0.007 & -0.065 & 5.3 & \\ 
		\hline
		\multirow{4}*{\textbf{4}}
		& D     & none  & unresolved &  828 & 0.409 [0.382,0.434] & -0.001 & -0.038 & 0 & traditional prescription\\ 
		& C     & none  & unresolved & 2012 & 0.644 [0.628,0.661] & -0.001 & -0.046 & 0 & traditional prescription\\ 
		& A     & none  & unresolved & 5784 & 0.490 [0.487,0.494] & -0.008 &  0.009 & 0 & traditional prescription\\ 
		& B     & none  & unresolved & 1729 & 0.403 [0.346,0.409] & -0.037 & -0.063 & 0 & traditional prescription\\ 
		\hline
	\end{tabular}
	\\
	$^1$ Region of the HRD being analysed: D = upper main sequence, $-1.5\leq\gmag\leq4.0$, $-1.0\leq(\gbp\!-\!\grp)_0\leq1.0$; C = upper main sequence plus evolved stars, $-1.5\leq\gmag\leq4.0$, $-1.0\leq(\gbp\!-\!\grp)_0\leq3.5$; B is the lower main sequence $4.0\leq\gmag\leq7.5$, $-1.0\leq(\gbp\!-\!\grp)_0\leq3.5$; A is the entire HRD from there above, that is $-1.5\leq\gmag\leq7.5$, $-1.0\leq(\gbp\!-\!\grp)_0\leq3.5$.
	\\
	$^2$ References for binary parameters: E06 = \citet{Eggleton2006}; MDS17 = \citet{moe17}. ``Resolved'' means that binary components with angular separation larger than the separation line in  Fig.~\ref{fig:resolved_bin} are counted as two distinct stars.
\end{table*}

\begin{figure*}
    \includegraphics[width=\textwidth]{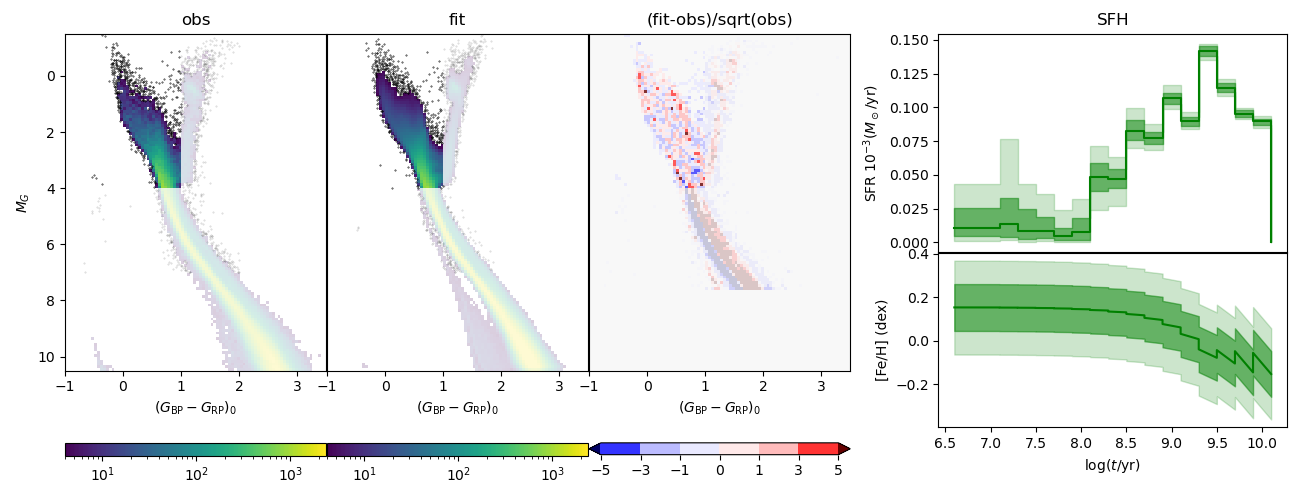}
    \caption{The fitting of the upper MS (region D), with limits $-1.5<\gmag<4$ and $-1.0<\gcolor<1$. The left panels show the Hess diagrams of the observations and the mean best-fitting model, in the same color scale, plus the map of residuals in approximate units of $\sigma$ -- that is, the result of $(M_k-O_k)/\sqrt{O_k}$. This latter is shown only where the comparison can be meaningful at all, i.e. for bins with more than 10 observed stars. The right panels shown the SFH, separated into two panels: the upper for the SFR$(t)$, the lower for the $\feh(t)$. For the SFR$(t)$, the central heavy green line marks the median value at each age, with two shaded areas marking the 68 and 95 per cent confidence intervals. For the $\feh(t)$, the central heavy green line mark the median value, while the two shaded areas mark the intervals that comprise 68 and 95 per cent of the metallicities effectively used in the mean best-fitting model, respectively.
    }
    \label{fig:fit_D}
\end{figure*}

\section{Discussion}
\label{sec:discuss}

Let us now discuss a few of the many different fittings made possible by our codes. The discussion is limited to the $\gmag\!<\!7.5$~mag interval, for which we can be sure the incompleteness is smaller than a few percent (cf. Section~\ref{sec:data}), and which includes the entire region of red giants and subgiants in the HRD. It is also large enough to contain a significant piece of the lower main sequence, which is very sensitive to the presence of unresolved binary systems. 
On the other hand, by limiting the HRD to this magnitude limit, we avoid having to discuss possible changes in the low-mass IMF \citep{sollima19, hallakoun20}, which we leave for subsequent papers. 

In the following, we comment the fittings in groups, or ``model families'' in Table~\ref{tab:chi2}, that represent common physical assumptions for the binaries.

\subsection{Analyses of different HRD sections with unresolved binaries}

Different regions of the HRD provide different constraints on the SFH, AMR, and binary fraction of the Solar Neighbourhood. In the following, we present the family of models labelled as \textbf{1} in Table~\ref{tab:chi2}. They make use of unresolved binaries computed with BinaPSE and with the \citet{Eggleton2006} distribution of initial masses and orbital parameters. All binaries are assumed to be unresolved. This family of models is \textit{not} our favoured one (as it will become clear in the next subsections), but it represents a well-accepted approach to model binaries in CMD-fitting works, and it illustrates some problems that are common to all models to be presented afterwards.

Figure~\ref{fig:fit_D} shows a fitting in which only the upper part of the MS (region D), with $\gmag<4$~mag and $(\gbp-\grp)_0<1$~mag, is included. This exercise essentially uses the simplest age indicator -- i.e. the number of stars still on the MS at every magnitude interval -- to derive the SFH. As can be appreciated, in this simple case we obtain an excellent fitting of the entire HRD region being considered.  The pattern of residuals is quite uniform, resembling the results obtained under ideal conditions (i.e. using mock catalogues) in the Appendix~\ref{sec:testing}. The SFH indicates a maximum SFR$(t)$ of $1.4\times10^{-4}$~\Msun$\mathrm{yr}^{-1}$ at ages of about 2 Gyr, with SFR$(t)$ being reduced both at older and younger ages. At ages younger than $10^8$~yr, we find  an upper limit of $\mathrm{SFR}(t)\lesssim4\times10^{-5}$~\Msun$\mathrm{yr}^{-1}$, which is not surprising for a sample that avoids the Galactic Plane. The binary fraction turns out to be well constrained, at $\fbin\simeq0.41$.

\begin{figure*}
    \includegraphics[width=\textwidth]{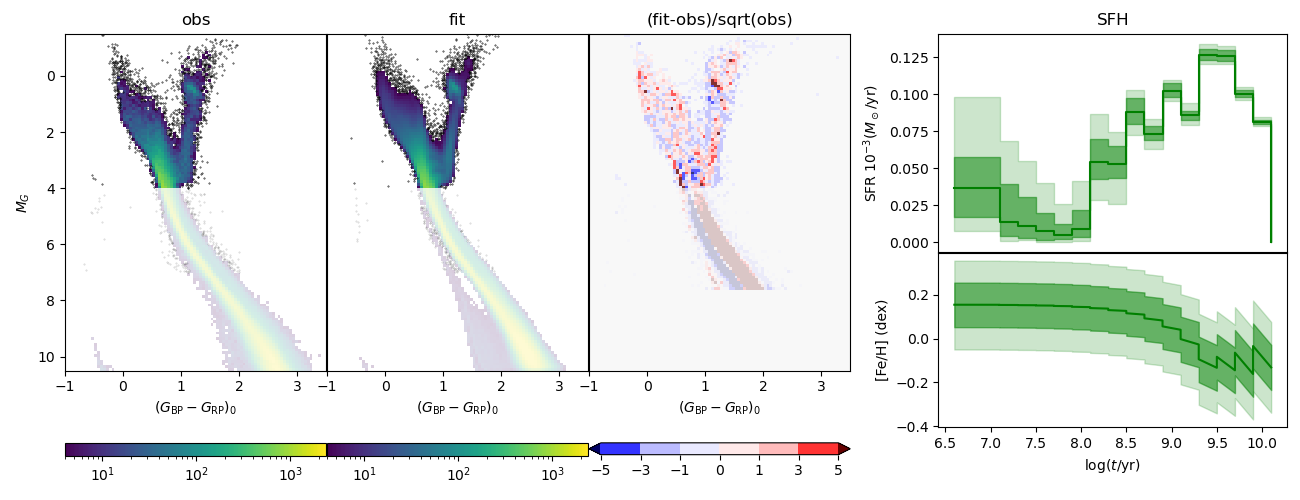}
    \caption{The same as Fig.~\ref{fig:fit_D} but now for case C, fitting the entire upper HRD with $-1.5<\gmag<4.0$ and $-1.0<\gcolor<3.5$.}
    \label{fig:fit_C}
\end{figure*}

Figure~\ref{fig:fit_C} then includes all the red giants and subgiants with $\gmag<4$~mag (region C) in the analysis. In comparison with the previous case, it can be immediately noticed that the fitting residuals increase in a few regions of the HRD, for instance close to the oldest MS turn-off ($\gmag\sim4$), and around the RC region. But over most of the HRD, the quality of this fit still looks acceptable. It is also evident that the general shape of the SFR$(t)$ is very much consistent with the one derived from the MS alone in Fig.~\ref{fig:fit_D}. The most evident differences compared to the previous case are in the AMR, which appears ``flattened'' for ages older than 2 Gyr, and in the binary fraction which increases to 0.63.

Despite these apparently-modest changes in the fittings, in passing from model D to C the changes in SFR$(t)$ and \fbin\ largely exceed those expected from the error bars of both best-fitting models. This result by itself indicates that the error bars derived from the MCMC -- despite being formally correct (see Appendix~\ref{sec:testing}) -- largely underestimate the true errors. At the very least, they do not include some source of systematic error, that forces the MCMC to pursue different solutions when selecting different areas of the HRD.

\begin{figure*}
    \includegraphics[width=\textwidth]{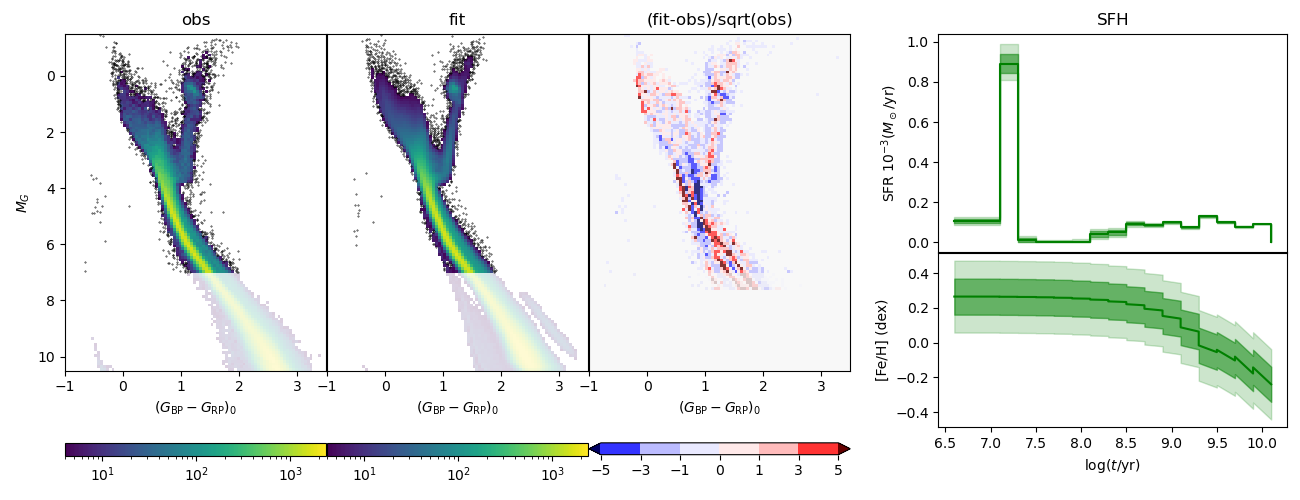}
    \caption{The same as Fig.~\ref{fig:fit_D} but now for case A, i.e. fitting the entire HRD above $\gmag=7.5$~mag.}
    \label{fig:fit_A}
\end{figure*}

Figure~\ref{fig:fit_A} shows the fitting of the entire HRD above $\gmag=7.5$~mag (region A in Table~\ref{tab:chi2}), i.e. now including a significant part of the lower MS. The changes in the results are dramatic. First, the residuals increase, especially close to the oldest MS turn-offs at $\gmag\simeq4$, and also along stripes on the lower MS. Regarding the SFR$(t)$, the most evident difference in case A is the appearance of a strong peak of young SFH at ages smaller that $\sim2\times10^7$~yr. Apart from this peak, the derived SFH is similar to those derived from cases C and D. Regarding the binary fraction, it turns out to be 0.42, i.e. intermediate between cases B and C. Another marked difference is that the presence of the low-MS also drives the solution to higher values of metallicity at young ages.

The strong peak in the very young SFH is probably an artefact, caused by the few observed stars being scattered at colours that are redder than the lower main sequence, and even redder than the sequence of equal-mass MS+MS binaries. In the partial models, the only stars that reach colours redder than the equal-mass MS+MS binaries are in short-lived pre-main sequence phases, present in the very young partial models. Therefore, the only way for our MCMC code to fit these few scattered red stars is to increase the SFR at very young ages -- within the limits allowed by the young massive stars present at the top of the observed MS. We regard this feature as indicative of the errors that can appear when a code is allowed to \textit{blindly} fit all stars present on a HRD. In our specific case, the ``too-red'' stars amount to about 500, and might be either artefacts, or unresolved triple systems (not included in our PMs).

\begin{figure}
    \includegraphics[width=\columnwidth]{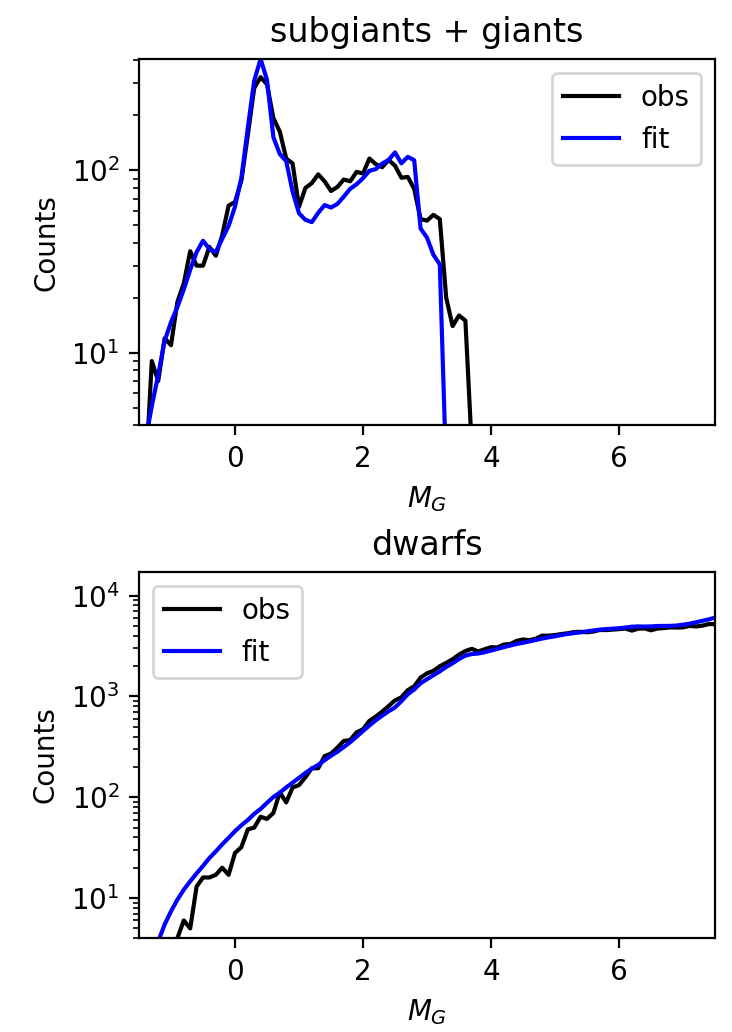}
    \caption{Luminosity functions for stars in model A compared to the observations, separately for subgiants+giants (top panel) and dwarfs (bottom panel). 
    }
    \label{fig:lfs}
\end{figure}

Another interesting fact is that the fitting of the entire HRD turns out to produce significant residuals close to the oldest MS turn-offs, at $\gmag\sim4$~mag. Such higher residuals were just slightly hinted at the previous case C above. It is worth noticing that these high bin-to-bin residuals are nearly cancelled if one looks at the residuals over larger fractions (or larger bins) of the HRD. This point is illustrated by the luminosity functions (LF) shown in Fig.~\ref{fig:lfs}, comparing data and best-fitting model for case A, separately for subgiants+giants and dwarfs. It can be appreciated that there is overall a good agreement between data and model LFs over the entire $-1.5<\gmag<7.5$ interval, with the notable exception of a small bump at $\gmag=2$~mag, which is present in the data but not in the model. This small discrepancy is probably related to the RGB bump which is slightly misplaced in PARSEC v1.2S tracks \citep[see][]{fu18}.

From the models presented so far, it is evident that case A provides the worst result, and the less reliable determination of the SFH in the Solar Neighbourhood. The reasons for this failure should probably be looked for in the details of the stellar models, which might not fit simultaneously the field stars observed across the very wide colour and magnitude intervals of Fig.~\ref{fig:fit_A} -- which also involve a wide range of masses, effective temperatures, and surface gravities. A careful testing of these models is required. It is worth remarking that the present isochrones were already used to fit the Gaia photometry of hundreds of open clusters \citep[see][]{babu, bossini19, monteiro20, dias21, medina21} with apparently good results, even when the photometry spans several magnitudes along the main sequence. But, taken individually, the open clusters contain too few evolved stars to provide a good quantitative test of our partial models.

\subsection{The fraction of unresolved binaries from the lower MS}
\label{sec:bin_lowMS}

\begin{figure*}
    \includegraphics[width=\textwidth]{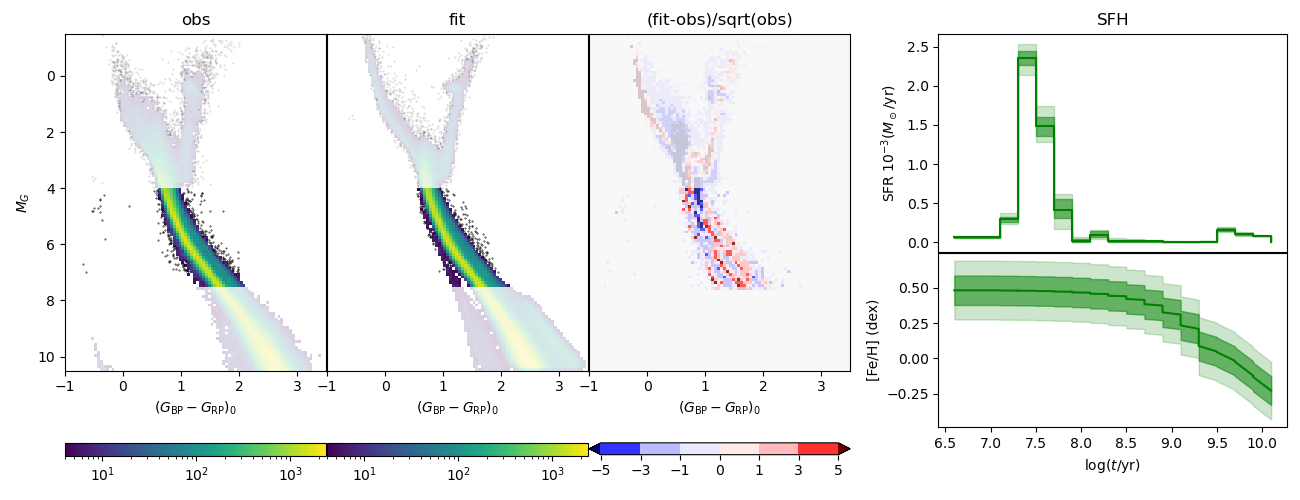}
    \caption{The same as Fig.~\ref{fig:fit_D} but now for case D, i.e. fitting only the lower MS. The results for the SFH have little relevance in this case, but are shown for the sake of completeness.
    } 
    \label{fig:fit_B}
\end{figure*}

The fittings presented above give contrasting results for the initial binary fraction, \fbin. Independent information about this fraction is provided by the lower main sequence (with $4<\gmag<7.5$, region B in Table~\ref{tab:chi2}). This HRD region presents the clearest manifestation of the presence of \textbf{unresolved} binaries, which are evident as the ``second MS'' observed to the right of the MS. This feature is known since \citet{haffner37} and evinces the formation of a relatively high fraction of near-mass binaries, with mass ratios in excess of $\sim\!0.7$. On the other hand, the bulk of stars in this HRD region are unevolved, and therefore it should provide little information about the SFH. Therefore, this exercise is mainly intended to constrain the binary fraction. 

The results are illustrated in Fig.~\ref{fig:fit_B}, which shows both the best-fit models and the derived SFH. As can be noticed, this kind of fitting clearly points to a binary fraction close to 0.23. 

Also in this case, the fitted solution favours a weird SFH in which there is a strong burst of very young star formation (with ages $\logtyr<8$) with a very high metallicity ($\feh\simeq0.5$~dex). This is likely the same problem that happened for the model A in Fig.~\ref{fig:fit_A}, which we regard as an artefact caused by the few scattered red stars. Moreover, the map of residuals indicates that this is far from being a perfect fit of the data, with some red/blue strips in the HRD indicating sub/overproduction of stars at the level of $\sim3\sigma$. This is in stark contrast with the level of agreement we are used to, while fitting the CMDs of external galaxies. One main culprit, in this case, is probably the extreme (and unusual) accuracy of the Gaia data we feed to the HRD-fitting algorithm. In external galaxies, accuracies better than a few hundredths of magnitude cannot be reached in the placement of stars at the bottom of the main sequence, even in the most favourable cases (namely for HST observations of the Magellanic Clouds, see \citealt{holtzman06} and \citealt{merica20} for examples). With Gaia DR2, this is possible, and the adoption of a similar resolution in the HRD provides very demanding constraints to the model fitting. Being our model prescriptions not perfect, the improved constraints result in higher-than-usual residuals in some places of the HRD. Irrespective of these problems, the main indication we get from this fitting is that about 1/4 of all low-MS stars are unresolved binaries. 

\subsection{Rough corrections from the recoverability of binaries in Gaia DR2}
\label{sec:unresolved}

The family of models \textbf{1}, just discussed, includes a common assumption in CMD-fitting codes: that all binaries are unresolved and hence always contribute to the HRD as single sources. While this assumption is valid for external galaxies analysed with similar methods, it is not for our nearby sample. Wide binaries are common in the Solar Neighbourhood and their catalogues were greatly expanded as a result of Gaia proper motions \citep[see][]{jimenez19, zavada20, hartman20}. For instance, we verify that the \citet{hartman20} catalogue contains 8775 wide binaries in our sampled volume, which means that twice as many ``apparently-single stars'' are included in our analysis. While this represents just a small fraction of our total sample, this might be just the tip of an iceberg that might be more clearly revealed after Gaia DR3.

On the other hand, the \citet{Eggleton2006} and \citet{moe17} prescriptions provide very broad distributions of initial semi-major axis $a$, with median values of the order of 1000~AU. This implies a large fraction of predicted binaries with separation in excess of a few arcsec, and hence well resolvable by Gaia.

\begin{figure}
	\includegraphics[width=\columnwidth]{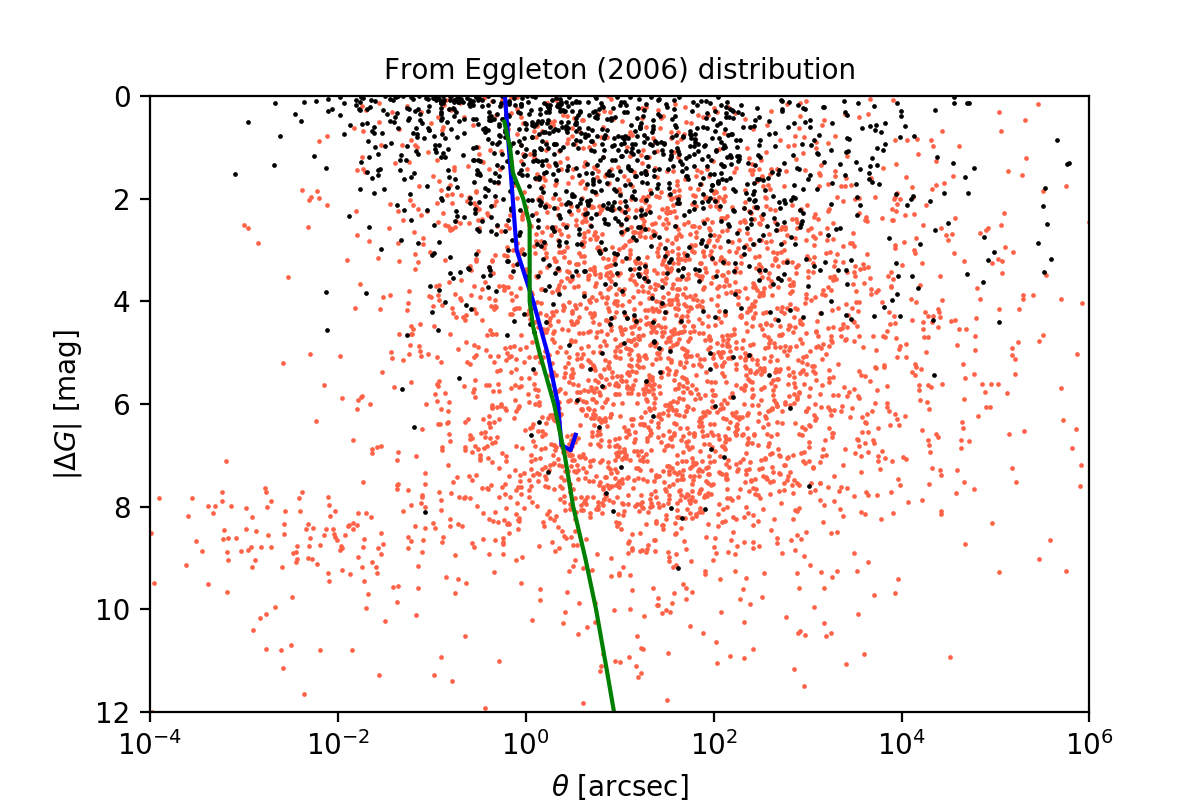}
	\includegraphics[width=\columnwidth]{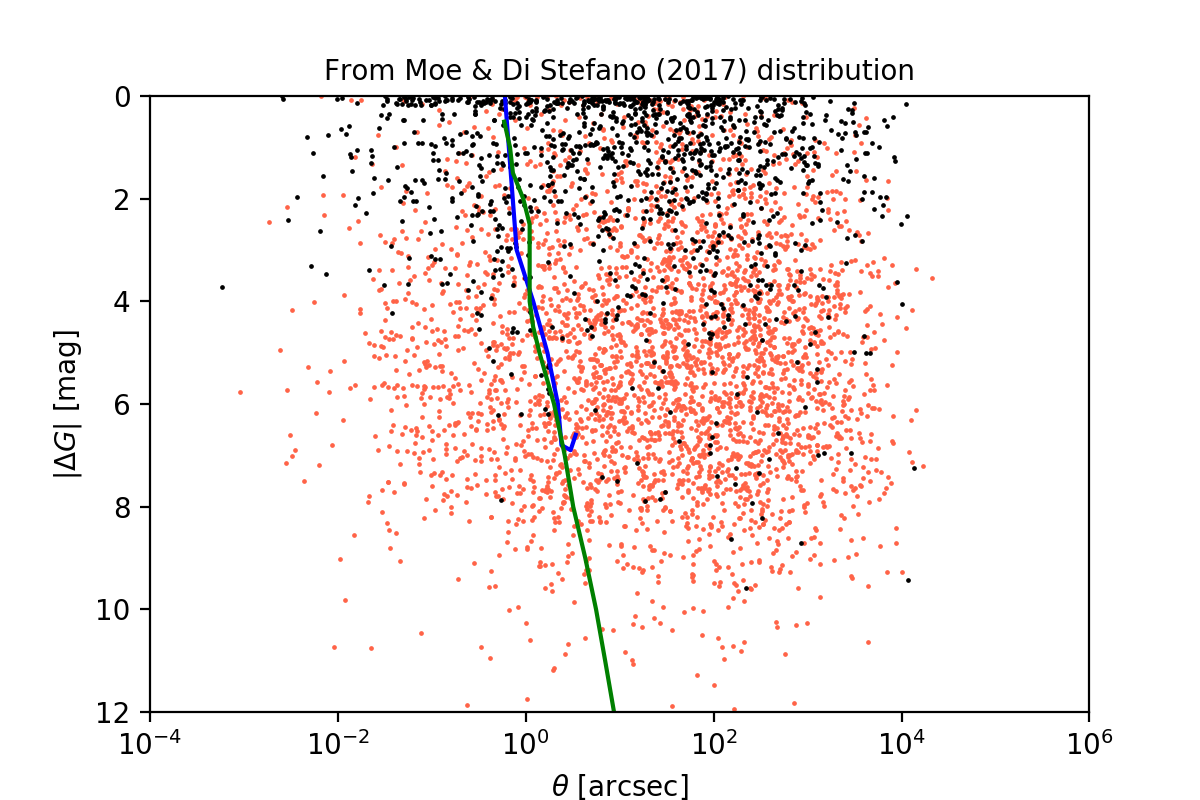}
    \caption{\textbf{Top panel}: A sample of simulated binaries within 200 pc (dots), plotting their angular separation versus magnitude contrast. Initial parameters are taken from the \citet{Eggleton2006} distribution. Black dots are binaries in which both components are bright enough to be observable as single stars; red dots are those in which only the brightest component is observable. The continuous blue line shows the 50\%-recoverabilty line defined by \citet{ziegler18}, while the green line is the same limit measured by \citet{brandeker19}. Binaries to their right have a high probability of being resolved in Gaia DR2. \textbf{Bottom panel}: The same for a simulation derived from \citet{moe17} initial distribution.
    }
    \label{fig:resolved_bin}
\end{figure}

To quantify the fraction of binaries that would be resolved in Gaia DR2, we use BinaPSE together with the \citet{Eggleton2006} distribution to simulate a uniform sample of binaries within 200~pc, made of stars of near-solar metallicity and with a constant SFR$(t)$, and subjected to the broad magnitude cuts described in Sect.~\ref{sec:data}. The present-day binary orbits are then attributed a random orientation and are ``observed'' at a random epoch. The top panel of  Fig.~\ref{fig:resolved_bin} shows their final distribution of the angular separation $\theta$, and the magnitude difference between the primary and secondary in the $G$ band, $|\Delta G|$. As can be appreciated, very broad distributions of these parameters are derived. 

The ``recoverability'' of binaries in Gaia DR2 has been measured by \citet{ziegler18} by means of a large imaging survey using adaptive optics, up to separations of 3.5$\arcsec$, and then by \citet{brandeker19} using Gaia DR2 data itself, for separations up to 12$\arcsec$. Their 50\%-recoverability limits are displayed in Fig.~\ref{fig:resolved_bin}. We adopt as a reference the curve drawn by using the point at $|\Delta G|=0$ from \citet{ziegler18} and then the \citet{brandeker19} curve for all other values. All binaries with a separation larger than given by this curve have a high probability of being included as two stars in Gaia DR2. In our simulations, 69 per cent of all binaries satisfy this condition, but for the large majority of them, only one of the components is bright enough to be included in our sample, implying that these binaries would anyway be counted as single stars. The fraction of our binary sample that satisfies the 50\%-recoverability limit and ensures both primary and secondary are counted, $\fresolved$, is of 0.18 for the total $\gmag<7.5$~mag sample (case A). For stars at the lower main sequence, with $\gmag>4$~mag (case B), $\fresolved$ increases to 0.73 -- mainly because for binaries in the lower MS the magnitude contrast $|\Delta G|$ is much more frequently limited to just a few magnitudes. For the upper part of the HRD ($\gmag<4$~mag, case C), instead, $|\Delta G|$ values become larger, and $\fresolved$ falls to 0.09.

Before proceeding, we recall that similar numbers are derived when we use the \citet{moe17} distribution of binary parameters, as illustrated in the bottom panel of Fig.~\ref{fig:resolved_bin}: the values of $\fresolved$ become 0.20, 0.80, and 0.11, for the A, B and C samples, respectively.

The distinction between resolved and unresolved binaries would imply significant corrections to the binary fraction. For every binary fitted in the HRD, there are $2\fresolved$ apparently-single stars that are, in reality, $\fresolved$ binaries. The approximate relation between the true binary fraction, $\fbin^\mathrm{true}$, and the value derived from unresolved binaries, $\fbin^\mathrm{app}$, is
\begin{equation}
\fbin^\mathrm{true} \simeq \frac{\fbin^\mathrm{app} + \fbin^\mathrm{true} \fresolved}{1-\fbin^\mathrm{true}\fresolved}  ,
\label{eq:correct0}
\end{equation}
which implies
\begin{equation}
\fbin^\mathrm{true} \simeq \frac{1-\fresolved - \sqrt{(\fresolved-1)^2-4\fresolved\fbin^\mathrm{app}}}{2\fresolved}  .
\label{eq:correct}
\end{equation}

Applying this correction, the binary fractions measured from cases A, B and C, are increased from 0.423 to 0.59, from 0.228 to 0.30, and from 0.635 to 0.99, respectively. It is evident that the consideration of the recoverability of the binaries can alter, significantly, the \fbin\ values found for the several sections of the HRD.

\subsection{Fully implementing unresolved and resolved binaries in the HRD fitting}

Therefore, we implement a more realistic simulation of the binaries, in the family of models referred to as \textbf{2} in Table~\ref{tab:chi2}. To prepare these models, the first step is to modify the preparation of the binary PMs: For every binary produced by BinaPSE, a random location, orientation and observation epoch is assigned so as to place it in Fig.~\ref{fig:resolved_bin}, hence defining whether it is counted as a single star, or as two stars in different HRD locations. All binary PMs are affected by this procedure.

The results of these HRD fittings are presented in the block \textbf{2} of Table~\ref{tab:chi2}. Compared to the previous family of models \textbf{1}, the fitting of the data is slightly worse (i.e.\ with larger values of $-\ln\mathcal{L}$), although the features that appear in the solutions are quite similar to the previous case. But the one aspect that strikes in all these solutions is that the \fbin\ values derived from the upper part of the HRD are quite large, i.e. between 0.8 and 1.0, whereas the lower MS indicates a value close to 0.4.

For completeness, we perform the same exercise using the \citet{moe17} distribution of initial binary parameters, producing the family of models in the block \textbf{3} of Table~\ref{tab:chi2}. Overall, the results are similar to those obtained with the \citet{Eggleton2006} distribution. In both cases, the \fbin\ values derived from the lower MS are nearly half the values derived from the analyses of upper parts of the HRD.

This particular result could be indicating two different things: either the binary fraction decreases with stellar mass much more rapidly than assumed in the \citet{Eggleton2006} and \citet{moe17} distributions, or the upper parts of the HRD, for some unknown reason, tend to be better fit with larger fractions of binaries than present in stellar populations. In order to explore this second possibility, we perform the family of models 2a, which simply fits the upper part of the HRD keeping the value of \fbin\ fixed at the 0.408 value inferred from the lower MS. For the cases D and C, the quality of the fitting results does not change much. For the case A, instead, the results are much worse, again reflecting the intrinsic difficulty of the stellar models to fit the entire HRD simultaneously. The impact of the choice of \fbin\ on the SFH will be discussed in Sect.~\ref{sec:SFH} below.

\subsection{On the usual assumption of unresolved binaries}
\label{sec:binfrac}

The model family \textbf{4} in Table~\ref{tab:chi2} presents the results of the HRD fitting using the classical prescription used in CMD fitting -- i.e.\ that all binaries are unresolved and have a flat distribution of mass ratios.  This prescription is adopted, for instance, by \citet{dolphin02}, \citet{rubele18}, \citet{ruiz20}, with only minor changes -- especially regarding the minimum value of mass ratio, which varies from author to author.
In our case, we have limited the mass ratios in the interval $0.7<q<1$, but extending this range would not change the situation much, since non-interacting unresolved binaries with smaller $q$ values are nearly identical to single stars. As we can observe in the table, this kind of model provides (a) nearly the same values of $\fbin$ (going from 0.40 to 0.64) for different parts of the HRD, and (b) fits nearly as good as the other cases already examined. In summary, nothing in our solutions indicates that these models are any worse than those obtained with more detailed prescriptions for the binaries. 

Indeed, the problem with these models is not in the quality of the HRD fitting they provide. It is simply that they do not offer the possibility of properly accounting for (and quantifying) the resolved and unresolved binaries in different samples. For instance, in our $d<200$~pc sample just a fraction of the binaries are in the form of unresolved systems (see Sect.~\ref{sec:unresolved}); therefore, it is not appropriate to assume that it has the same fraction of unresolved binaries that were used to fit, successfully, the CMDs of nearby galaxies. And considering the very wide distributions of orbital separations provided by \citet{Eggleton2006} and \citet{moe17} prescriptions (Fig.~\ref{fig:resolved_bin}), this problem should affect, in a significant way, even samples built for distances ten times larger than our $200$-pc limit. The problem only disappears at distances of \textit{tens of kiloparsecs}, as for instance in the Magellanic Clouds, where the fraction of resolved binaries becomes negligible.

\subsection{Indications about the fraction of close binaries}

Hot subdwarfs derive only from binaries which have the chance of exchanging mass before the onset of helium core-burning, as illustrated in the example of Sect.~\ref{sec:hestars}. Therefore they specifically sample the fraction of binaries formed in close orbits. Moreover, as indicated in Table~\ref{tab:PMs}, the formation of hot subdwarfs peaks at ages close to 1~Gyr, i.e. just before the onset of extended RGBs in stellar populations. These ages correspond to turn-off masses of about 2~\Msun, and therefore to stars observed at $\gmag\lesssim2$~mag while on their MS. These magnitude intervals are well-sampled and well-fit in our HRDs. 

Therefore, it was to be expected that our best-fit models would reproduce the observed numbers of hot subdwarfs, but this is not exactly the case. There are 17 hot subdwarfs in the observations (see Fig.~\ref{fig:bighrd}). Our best models for cases D and C predict between 19 and 22 of them (see Table~\ref{tab:chi2}, for model families 2 and 3), but these cases are likely overestimating \fbin\ by a factor of about 2 (see Sect.~\ref{sec:bin_lowMS}). The alternative set of models 2a, build for the $\fbin=0.408$ value inferred from the lower MS, indicates an expected number of 8.3 hotdwarfs, i.e. a factor of 2 less than observed.

Considering the low-number statistics, the discrepancy is not dramatic, but it is slightly worrying -- indeed, the probability of observing $\geq17$ out of expected 8.3 stars is just 0.5 per cent. In the case this deficit is confirmed by future studies including larger subsamples of Gaia data, it is worth recalling that: 
\begin{itemize}
    \item The original BSE code generates about 2.4 times \textit{less} subdwarfs than BinaPSE, for the same SFH (last row of Table~\ref{tab:PMs}).
    \item Only stars with initial separation closer than $\sim\!1$~AU produce hot subdwarfs, therefore this is the range for the possible revisions in the distribution of initial binary parameters in \citet{Eggleton2006} and \citet{moe17} prescriptions.
    \item The number of predicted hot subdwarfs depends on a series of other parameters in the binary evolution code, in addition to the distribution of initial separations. In particular, the production of hot-subdwarfs is expected to increase with the common-envelope ejection efficiency and the critical mass ratio for dynamically unstable mass transfer \citep[see][]{heber09, han02, han03}.
\end{itemize}
This probably means that these parameters have to be recalibrated in BinaPSE, in order to produce about twice more hot subdwarfs. The Gaia data represents the ideal sample to face this problem in the near future. As the production of hot subdwarfs changes with age (Table~\ref{tab:PMs}), this recalibration goes hand in hand with the determination of the SFH in the Solar Neighbourhood.

\begin{figure*}
	\includegraphics[width=\textwidth]{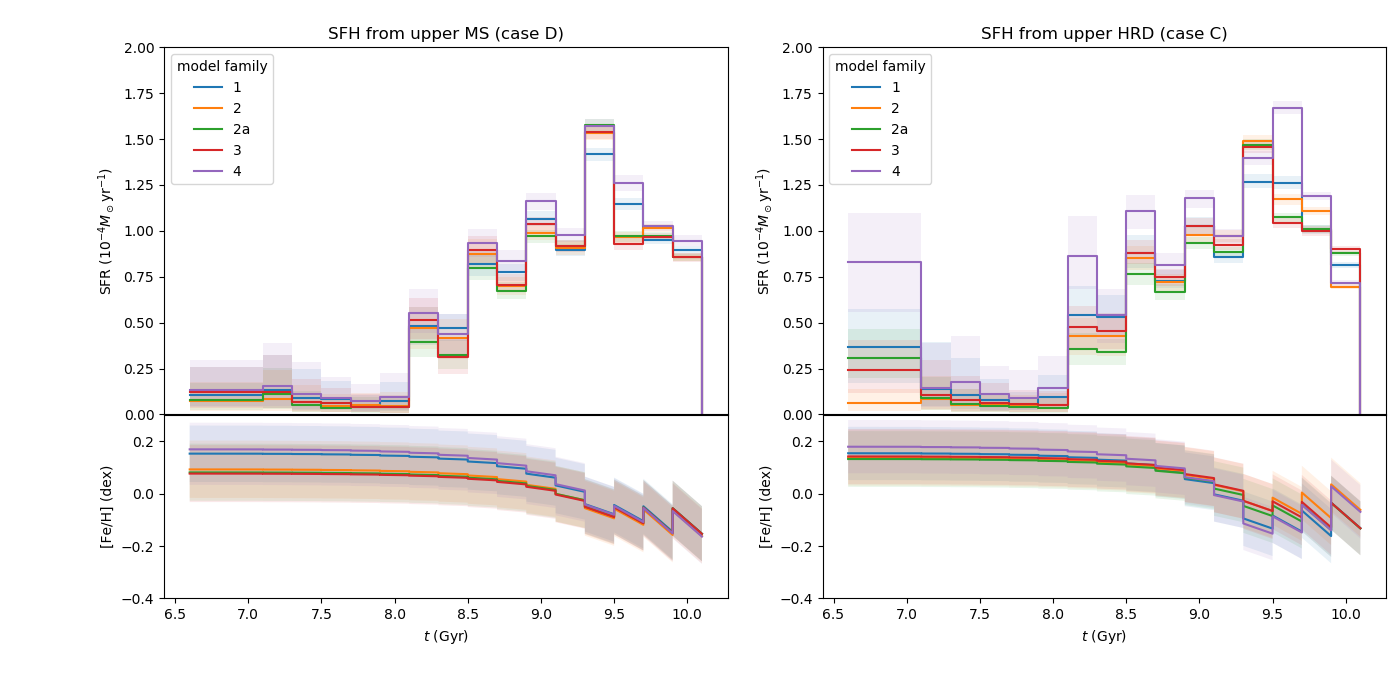}
    \caption{
    Several of the SFHs derived in this work, for models introduced in Table~\ref{tab:chi2}. Left panels correspond to the results obtained by fitting the upper MS, the right ones to the fitting of the entire uper HRD. In both cases the upper panel shows the SFR$(t)$, with solid lines corresponding to the median solutions, and light-shaded areas illustrating the 16\% to 84\% confidence interval. The bottom panel shows the median AMR, \feh$(t)$, for the same models, together with the metallicity interval that spans 16\% to 84\% of the values adopted at every age.}
    \label{fig:sfh_comp}
\end{figure*}

\subsection{The SFH in the Solar Neighbourhood}
\label{sec:SFH}

In this work, we concentrate on the analysis of the basic building blocks and assumptions of CMD (or HRD) fitting. But assuming that our best models (cases C and D) are good enough representations of reality, we also derive a potentially important constraint to Galaxy evolution models: the SFH of the Solar Neighbourhood.

The left panel of Fig.~\ref{fig:sfh_comp} shows several of the solutions we find along this work, limited to the upper MS (case D in Table~\ref{tab:chi2}), which provide the smallest residuals. As can be appreciated, all these models produce qualitatively similar SFR$(t)$, with the maximum being in the age bin between 2--3.2 Gyr and with values between $1.25\times10^{-4}$ and $1.5\times10^{-4}\Msun\mathrm{yr}^{-1}$. Other maximums are found at the age bins 0.8--1.2 Gyr and 0.32--0.5 Gyr. Despite the limited age resolution we have adopted for older ages, the SFR$(t)$ clearly decreases up to the 12.6~Gyr limit of our determination, reaching values of $\sim\!0.85\times10^{-4}\Msun\mathrm{yr}^{-1}$.

The mean $\feh(t)$, instead, appears generally decreasing by about 0.3~dex between young and old ages, in all cases, although following slightly different median curves. If we consider the Gaussian dispersion of $\sigma=0.1$~dex adopted to build the PMs at every age, there is a great degree of superposition between the metallicity distributions represented by these different cases.

The right panel of Fig.~\ref{fig:sfh_comp} presents the results from the analyses of the entire upper HRD (case C in Table~\ref{tab:chi2}). The quantitative differences between the different SFR$(t)$ become slightly larger in this case, but anyway essentially the same kind of solutions are found as in previous case D. The most discrepant results, in this case, come from models in family 4, in which all binaries are non resolved and have a flat distribution of mass ratios: these particular solutions present a SFR$(t)$ peaking at slightly older ages than in the other cases, that is in the 3.2--5 Gyr age bin.

Overall, these plots reveal that the prescription adopted for the binaries has a modest impact on the final derivation of the SFH. Based on Fig.~\ref{fig:sfh_comp}, we estimate that the uncertainty on the binary prescription alone adds an uncertainty of about 20 per cent on the values of SFR$(t)$ at any age. The impact on the AMR seems also quite modest.

How our SFHs compare to other ones recently derived from Gaia? Any direct comparison with other authors is complicated by the very different samples they analysed -- in addition to the different methods and stellar models they adopted. Therefore we refrain from a doing detailed comparison. But it is worth to mention that:
\begin{itemize}
    \item We do not find the marked increase in the SFR$(t)$ for ages older than about 10~Gyr that can be appreciated in the solution by \citet{ruiz20}. Their sample refers to a wider volume, extending up to 2~kpc from the Sun, therefore higher fractions of old stars (and lower metallicities) are expected from their sampling of larger sections of the thick disk and halo. What comes as a surprise is that a similar increase in the star formation rate of old stars (with ages $\ga8$~Gyr in this case) can also be inferred from the \citet{alzate20} ``age-metallicity distributions'', which were derived from a sample extending to just 100~pc. 
    \item Our solutions do not provide hints of the SFR$(t)$ peaks detected by \citet{ruiz20} at ages 1.9 and 5.7 Gyr, while we might agree that there is a SFR$(t)$ peak at 1~Gyr. The absence of the older peaks in our case might be simply caused by the the smaller sample and the coarse age resolution we adopted in our method. It is notable that our solution for case 2a produces a SFR$(t)$ peak in the 3.2--5 Gyr age bin, which could be related to the 5.7~Gyr peak found by \citet{ruiz20} using similar assumptions for the binaries. However, in the absence of a detailed reanalyses of both samples using the same methods, this could be regarded as a coincidence.  
\end{itemize}


\section{Summary and conclusions}

In this work, we develop methods to simulate stellar populations including detailed prescriptions for binaries, and apply them to the fitting of the high-quality HRD from Gaia DR2.
We take special care to limit the analyses to an almost-complete sample little affected by the known errors and artefacts in the Gaia catalog; these considerations lead us to limit the sample to a maximum distance of 200~pc and to exclude the $|b|<25^\circ$ region. We put a particular effort into verifying if the populations of binaries can be well fit with present prescriptions. Our analysis stands on our refined binary evolution code BinaPSE, which is able to produce the main features caused by binaries in the Gaia HRD -- namely the sequence of near-equal mass binaries along the lower main sequence, and the isolated group of hot subdwarfs. Some indications from this work are:
\begin{itemize}
    \item Attempts to fit the entire HRD from Gaia DR2 within 200 pc, using its full colour-magnitude resolution from the lower MS to the upper MS, may fail to provide good solutions. Significant residuals appear probably as a result of discrepancies in the modelling of stars across a wide range of masses, effective temperatures, surface gravities, and metallicity. Zero-point errors in the Gaia photometry may also be playing a role. Until these problems are not completely characterised and their related parameters recalibrated (hopefully with the help of star clusters in the Gaia database), we recommend the fitting of separated and smaller sections of the Gaia HRD.
    \item Attempts to fit the initial fraction of mass formed in binaries, $\fbin$, will in general provide different results depending on the section of the HRD being analysed. This is illustrated by the case in which all binaries are assumed unresolved, which leads to \fbin\ values increasing from $\sim0.2$ to 0.6 as we go from the lower to the upper MS (assuming the \citet{Eggleton2006} prescription for the binary parameters together with the \citet{kroupa02} IMF).
    \item The problem remains when we use a more realistic prescription in which binaries are considered either as resolved or unresolved depending on their instantaneous angular separation compared to the 50\%-recoverability line measured for Gaia DR2. We have verified this for models adopting both the \citet{Eggleton2006} and the \citet{moe17} distributions of initial binary parameters, finding that the lower MS favours \fbin\ values close to 0.4, while the best-fit solutions of the upper-HRD favour values closer to 0.8 or 0.9. Solving this problem probably requires a more detailed analysis of the distributions of the adopted binary parameters, compared to observed samples.
    \item Anyway, when we adopt the \fbin\ derived from the lower MS -- which is certainly very sensitive to the fraction of unresolved binaries -- the fitting of the upper HRD is not dramatically worse than in the case where \fbin\ is also fitted. Overall, the star counts across the upper parts of the HRD are less affected by the presence of binaries, while extremely sensitive to the SFH.
    \item Our models provide the expected numbers of hot subdwarfs, produced from the evolution of close binaries especially at ages close to 1~Gyr. 17 such stars are present in our Gaia sample, while predicted numbers are 20--22 when both the SFH and $\fbin$ are fitted using the upper part of the HRD. Predicted numbers of hot subdwarfs decrease to 8.3 when we fix the \fbin\ value at $\simeq0.4$ as suggested by the lower MS. In any case, we regard reproducing this number within a factor of 2, without any additional tuning of model parameters, as a very promising start.
    \item In addition to the binary fraction, the method returns the SFH that best fits the HRD, in units of $\Msun\mathrm{yr}^{-1}$. The results depend moderately on the assumptions adopted for the binaries, and on other model aspects such as the section of the HRD and of the age-metallicity space being explored by the fitting algorithms. Anyway, the results clearly indicate a SFR$(t)$ peaking at ages close to 2~Gyr. Representative values for the SFR$(t)$ vary between the $\sim\!1.5\times10^{-4}\Msun\mathrm{yr}^{-1}$ at the peak, and $\sim\!0.8\times10^{-4}\Msun\mathrm{yr}^{-1}$ at old ages.
\end{itemize}
In summary, this work presents an independent analysis of the Gaia DR2 HRD and its SFH, producing generally good solutions in the upper part of the HRD, but also introducing a series of problems and questions that might be faced as future data releases from Gaia provide better data and larger samples. Overall, we are very pleased to find that \textit{the general goal of accurately fitting the star counts across the Gaia HRD, including their binary sequences, does not seem beyond reach}. 

Progress into improving the fitting of these HRDs certainly passes through a more detailed check of the binary models produced by our codes, and of their recoverability in Gaia. It is worth remarking that BinaPSE produces not only detailed information about the fraction of binaries across the HRD,
but also the distribution of other observables like the radial velocities, angular separations and magnitude contrast (as in Fig.~\ref{fig:resolved_bin}), proper motions, and light curves for eclipsing binaries.
In forthcoming works, we will add comparisons between the predictions of our TRILEGAL+BinaPSE codes with the wealth of data regarding binaries (including their white dwarf remnants) provided by other wide-field photometric and spectroscopic surveys, in the pursuit of more satisfactory prescriptions for modelling single and binary populations in the fields of nearby resolved galaxies. This work is a necessary step to prepare for the revolution in binary data expected from Gaia DR3, and from ambitious imminent surveys like the Vera Rubin Observatory Legacy Survey of Space and Time \citep{lsst} and the Sloan Digital Sky Survey V \citep{sdssv}.

\section*{Data Availability}
The data underlying this article were accessed from the ESA Gaia archive (\url{https://gea.esac.esa.int/archive/}). The best-fitting models generated in this research will be shared on reasonable request to the corresponding author.

\section*{Acknowledgements}
We thank Onno Pols, Omar Benvenuto, and the anonymous referee for the useful comments, Eduardo Balbinot, Meredith Durbin and Giada Pastorelli for the programming tips, and the organizers of the COST MW-Gaia WG2/WG3 for an inspiring workshop in Zagreb.
L.G.\ acknowledges partial funding from ``INAF mainstreams 1.05.01.86.20". P.M., P.D.T, L.G.,G.C. acknowledge the financial support from the European Union (ERC Consolidator Grant project STARKEY, no.\ 615604).
A.B.\ acknowledges PRIN MIUR 2017 prot.\ 20173ML3WW. Y.C.\ acknowledges NSFC funding No.\ 12003001. G.C.\ acknowledges financial support from the European Research Council for the ERC Consolidator grant DEMOBLACK, under contract no.\ 770017.



\bibliographystyle{mnras}
\input{gaia_cmd.bbl}

\appendix

\section{Examples of binary evolution with BINAPSE}
\label{sec:examples}

In this Section we describe in detail some examples of binary evolution that involve the main evolutionary channels and interaction process in binary systems, namely mass transfer, CE and mergers. Then, we present simulations of simple stellar populations (SSPs) with initial binary fraction equal to 1.0, metallicity $Z=0.007$, distance 1 kpc, initial mass $10^5\,M_\odot$ and ages 0.5, 1.0, 3.0, and 12.0 Gyr. All these SSPs have been simulated with TRILEGAL according to the distributions described in Section \ref{sec:probdist} and by evolving binaries in three different ways: as if they were non-interacting, with the BinaPSE code and with the original BSE code.

\subsection{The evolution of an Algol system}
\label{sec:algol}
\begin{figure}
    \centering
    \includegraphics[width=\columnwidth]{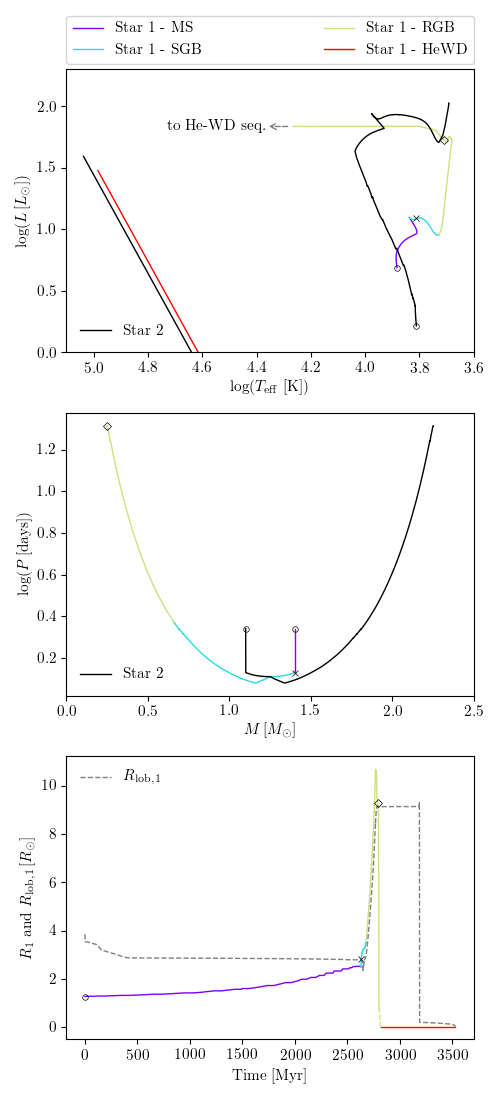}
    \caption{The evolution of the Algol binary system described in Section \ref{sec:algol}. The evolutionary path of star 1, the initially more massive one, is color coded according as indicated in the upper legend. Black solid lines represent the evolutionary path of star 2, the secondary. Initial positions are marked with open circles, the RLOF onset and end with crosses and open diamonds respectively. The upper panel shows the HRD, cut at $\log(L/L_\odot)=0$. The middle panel shows the dependency between the period and the stellar masses until the end of the mass transfer phase. The lower panel shows the evolution of star 1 radius and how it is affected by the Roche lobe radius.}
    \label{fig:ex1binapse}
\end{figure}

Consider a binary system with the following initial parameters: $m_{\mathrm{i},1}=1.4\,M_\odot$, $m_{\mathrm{i},2}=1.1\,M_\odot$, $P=2.2$ days, $e=0.5$ and metallicity $Z=0.007$. The evolution of this system is illustrated in the three of panels of Fig. \ref{fig:ex1binapse}. We call primary the initially more massive star, i.e. star 1. The evolutionary path of star 1 is color coded according to the evolutionary phase (see upper legend for details). The black solid lines, instead, represent the evolutionary path of star 2, the secondary. The two stars start their evolution on the zero-age main sequence (ZAMS). We marked the initial positions with open circles. As we can see in the HRD of the upper panel, star 1 evolves faster and it is able to complete the MS phase before filling its Roche lobe during the sub-giant phase. At this moment, marked with a cross in Fig. \ref{fig:ex1binapse}, orbit circularization and corotation of the stars with the orbit have already been achieved because of tidal interactions. The latter are also responsible for the orbital shrinkage before the RLOF which corresponds to the initial decrease of $P$ visible in the middle panel. Then, star 1 loses mass via RLOF for all the remaining part of the sub-giant phase and also in the red-giant phase. In the middle panel we clearly see how the hydrogen-rich material lost by the primary is accreted by the MS secondary during the mass transfer process while the binary period increases from 1.3 to over 20 days. The separation, $a$, analogously increases from 6.0 to over 42 $R_\odot$. In the lower panel we can easily identify the instants of RLOF onset and end as the intersections between the primary radius, $R_1$, and the radius of its Roche lobe, $R_{\mathrm{lob},1}$. Moreover, in the same panel, we can note that $R_1$ is slightly above $R_{\mathrm{lob},1}$ during the mass transfer phase. This is typical of the so-called thermal mass transfer: it is unstable on a thermal timescale and the donor does not remain in thermal equilibrium so it contracts and just fills the Roche lobe. The accretion of the material lost by the primary envelope allows the secondary to move upward along the ZAMS until the mass transfer terminates. When the mass transfer terminates (open diamond points in Fig. \ref{fig:ex1binapse}) the primary has lost about 1.15 $M_\odot$ and it has not completed the red-giant phase yet. The remaining envelope is not massive enough for the star to reach the critical mass for helium ignition so it leaves the Hayashi line and becomes a 0.25 $M_\odot$ He-WD. On the other hand, the secondary evolves now as a 2.25 $M_\odot$ ZAMS star. 

At this point the binary can be defined as an Algol system, i.e. a binary where the secondary has become more massive than the primary. The secondary overfills its Roche lobe during the red-giant phase but this time the mass transfer is of a different nature. The donor is a red giant as in the first case, but the ratio between its mass and the companion exceeds a critical value, $q_{\mathrm{crit}}$, defined as in \cite{hurley02}. Under these conditions the radius of the donor increases faster than the Roche lobe radius and this fact makes mass transfer unstable on a dynamical timescale. For this reason we refer to this process as dynamical mass transfer. The consequence of dynamical mass transfer is a CE phase that, in this case, leads to the expulsion of the giant envelope from the system. The result is a double He-WD binary system because, again, the 0.32 $M_\odot$ core of the giant star was not massive enough to ignite helium. After the CE phase the separation is about 0.6 $R_\odot$, but the orbit shrinks because of gravitational radiation and finally the two He-WDs merge. The temperature reached after the merging is assumed to be high enough to ignite the triple-$\alpha$ reaction. The energy released by the nuclear runaway is greater than the binding energy and the binary is completely destroyed. The total lifetime of the binary is about 3.5 Gyr.

\subsection{The formation of naked helium stars}
\label{sec:hestars}
\begin{figure}
    \centering
    \includegraphics[width=\columnwidth]{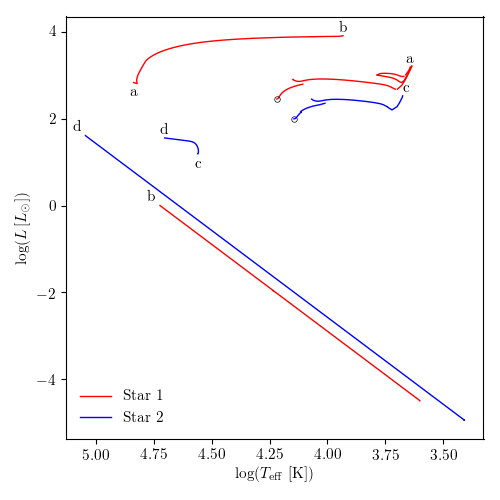}
    \caption{The evolution in the HRD of the binary system described in Section \ref{sec:hestars}. The evolutionary path of star 1, the initially more massive one, and star 2 are the red and blue solid lines respectively. Open circles mark the initial positions on the ZAMS and equal letters link the positions of the stars before and after a CE phase or any sudden evolutionary transformation. The alphabetical order of letters reflects the chronological order of such events.}
    \label{fig:ex2binapse}
\end{figure}

Naked helium stars form when, in a binary system, the hydrogen envelope of one of the two components is removed because of a mass transfer event or even expelled from the system after a CE phase. In this example we see two possible evolutionary channels leading to the formation of a naked helium star. We consider a binary system with $m_{\mathrm{i},1}=4.0\,M_\odot$, $m_{\mathrm{i},2}=3.0\,M_\odot$, $P=0.5$ yr, $e=0.5$ and metallicity $Z=0.007$. The evolution of the two stars in the HRD is shown in Fig.~\ref{fig:ex2binapse}: the primary path is in red, the secondary path in blue. Open circles mark the initial positions on the ZAMS and equal letters link the positions of the stars before and after a CE phase or any sudden evolutionary transformation. The alphabetical order of letters reflects the chronological order of such events. The primary is able to reach the early asymptotic giant branch phase (E-AGB) before filling its Roche lobe (position `a'). At that moment the secondary is still on the MS and a CE phase begins. The hydrogen envelope of the primary is lost by the system leaving a so-called helium Hertzsprung-gap star (HeHG star) of $1.06$ $M_\odot$ located at $(\log(T_{\mathrm{eff}}/\mathrm{K}),\,\log(L/L_\odot)) \simeq (4.8,\,2.8)$. This is a naked helium star with a CO core as massive as the progenitor core mass, that is burning He in a shell. It evolves very quickly towards lower temperatures and after a few Myrs it has lost the helium envelope by winds (position `b') leaving a $0.66$ $M_\odot$ CO-WD. HeHG stars more massive than this experience mass loss rates so high that allow us to classify them as Wolf-Rayet stars \citep{Woosley2019} and their peculiar composition makes them suitable candidates as progenitors of Supernovae Ib and Ic \citep{pols97,Dessart2020}. In the subsequent evolution, when the secondary reaches position `c' during the RG phase, a second CE phase begins. Again the hydrogen envelope of the giant is removed and lost by the system. Therefore, the secondary becomes a $0.47$ $M_\odot$ helium main sequence star (HeMS star), i.e. a naked helium star that is burning helium in the core. The HeMS phase lasts more than the HeHG phase in general, but even so it is quite fast (about $110$ Myr in this case). HeMS stars 
form the hot-subdwarfs group that is very well populated in Gaia HRDs. The mass of the HeMS star is around $0.47\,M_\odot$, so it is not able to produce a significant carbon core, it avoids the HeHG phase and becomes a $0.33$ $M_\odot$ He-WD. The final system is a double-degenerate system with a CO-WD and a He-WD that survives at least until $12$ Gyr when the simulation is stopped.

\subsection{The birth of a new E-AGB star}
\label{sec:eagbstar}
\begin{figure}
    \centering
    \includegraphics[width=\columnwidth]{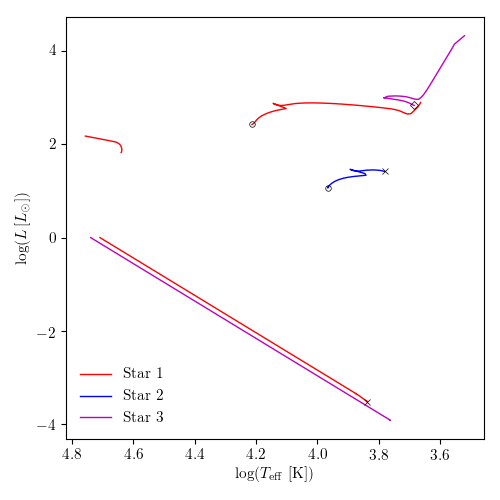}
    \caption{The evolution in the HRD of the binary system described in Section \ref{sec:eagbstar}. The evolutionary path of star 1, the initially more massive one, and star 2 are the red and blue solid lines respectively. Open circles mark the initial positions on the ZAMS. The crosses and the open diamond points mark the positions of the binary components just before the coalescence and the product of the merging respectively.}
    \label{fig:ex3binapse}
\end{figure}

We now consider a binary system with $m_{\mathrm{i},1}=3.93\,M_\odot$, $m_{\mathrm{i},2}=1.71\,M_\odot$, $P=0.35$ yr, $e=0.66$ and metallicity $Z=0.007$. The peculiarity of this system with respect to the previous examples consists in a merging between a CO-WD and a HG star whose outcome is a new E-AGB star. The evolution of the two stars in the HRD is shown in Fig. \ref{fig:ex3binapse}: the primary path is in red, the secondary path in blue and the new E-AGB star path is in magenta. Open circles mark the initial positions on the ZAMS. The primary evolves until it fills its Roche lobe during the red-giant phase with a CE formation. The hydrogen envelope of the primary is lost by the system leaving a 0.68 $M_\odot$ HeMS star which evolves into a 0.60 $M_\odot$ CO-WD. The secondary, still close to the initial state, completes the MS phase and fills its Roche lobe during the HG phase. Thermal mass transfer allows the formation of an accretion disk around the CO-WD. The CO-WD accretes part of the hydrogen that is burnt into helium on the WD surface. Nova explosions occur and are responsible for the ejection of a fraction of the accreted material. At this point the binary system can be classified as a cataclismic variable (CV). In our case the CO-WD mass does not grow enough and the supernova explosion does not take place. On the other hand, the accretion rate, $\dot{m}_1$, constantly increases. At first it reaches the value of $1.03\cdot 10^{-7}\,M_\odot\,\mathrm{yr}^{-1}$, the novae sequence interrupts and a steady accretion phase begins. During this phase the binary is a supersoft X-ray source. Then $\dot{m}_1$ increases above $2.71\cdot 10^{-7}\,M_\odot\,\mathrm{yr}^{-1}$ and the CO-WD should become a TP-AGB star. However, immediately after the transformation of the WD into a TP-AGB star, the system passes through another CE phase and the final coalescence of the two components. The outcome of the merging is a new 1.86 $M_\odot$ E-AGB star. In Fig. \ref{fig:ex3binapse}, the progenitors are marked with crosses and the position of the new E-AGB star is marked with and open diamond. The E-AGB star evolves as a single star, it passes through the TP-AGB phase and finally becomes a 0.69 $M_\odot$ CO-WD. Details about the formation of merging products can be found in \cite{hurley02}. The simulation is stopped at 6 Gyr.

\subsection{Binary evolution: comparisons with BSE}
\label{sec:BSEcomp}
The original BSE code is publicly available and anyone can compare the evolution of the binary systems described above with the BSE output. The BinaPSE and BSE results are qualitatively similar, i.e. they predict the same sequence of binary interactions and evolutionary phases and the same final products. However, quantitative differences are especially visible in the HRD and in the evolutionary timescales. As an example, we show in Fig. \ref{fig:ex3binapse} the evolution of the binary system of Sec. \ref{sec:algol} simulated with BSE. We superimposed in the HRD (upper panel) the evolutionary tracks predicted with BinaPSE (dashed lines). In the lower panel we compare the radius of the primary star in the two simulations and we note that mass transfer begins about 120 Myr later in the BSE simulation and it lasts 150 Myr more. As a consequence of this and of the different evolutionary timescales of the secondary, the final disruption of the system is postponed by 350 Myr. Finally, we remind that BSE analytic formulae don't take into account the first dredge-up. Therefore, BSE is not able to reproduce the RGB bump of stellar populations.

\begin{figure}
    \centering
    \includegraphics[width=\columnwidth]{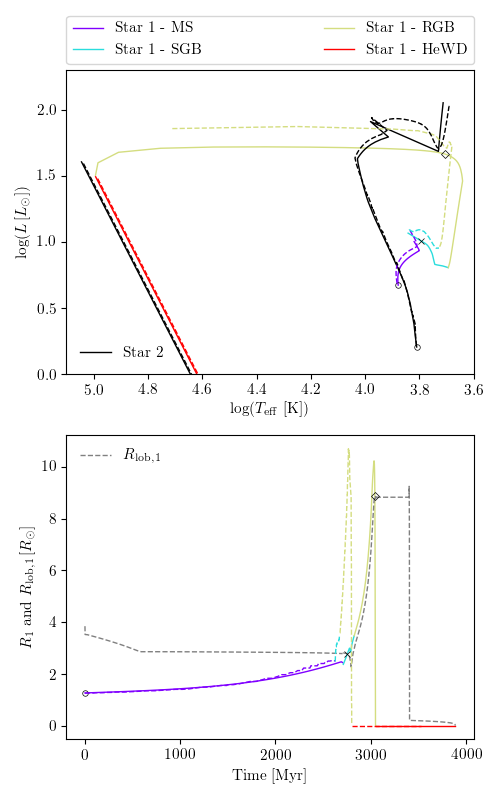}
    \caption{The evolution in the HRD of the binary system described in Section \ref{sec:algol} as modelled by BSE in comparison with predictions by BinaPSE. BSE evolutionary tracks follow the same notations of Fig. \ref{fig:ex1binapse}. Coloured dashed lines indicate star 1 evolution with BinaPSE. The black dashed line in the upper panel is star 2 evolution with BinaPSE.}
    \label{fig:ex1bse}
\end{figure}

It is worth reminding that the PARSEC tracks result from a decades-long work of fine-tuning of stellar parameters such as the mixing-length parameter, the helium-to metal enrichment ratio, and the efficiency of envelope and core overshooting. The adopted parameters allow these tracks to reproduce, in an acceptable way, a wide variety of observations of stellar aggregates dominated by single stars. Moreover, the algorithms we use to interpolate PARSEC tracks as a function of mass and metallicity, are extensively tested and are the same that are used to produce the widely-used PARSEC isochrones. Therefore, it is reasonable to expect that these tracks represent an improvement over the fitting relations originally adopted in BSE.

\subsection{Simulation of SSPs with BinaPSE and BSE}
\label{sec:binapseXbse}

\begin{figure*}
    \centering
    \includegraphics[width=0.88\textwidth]{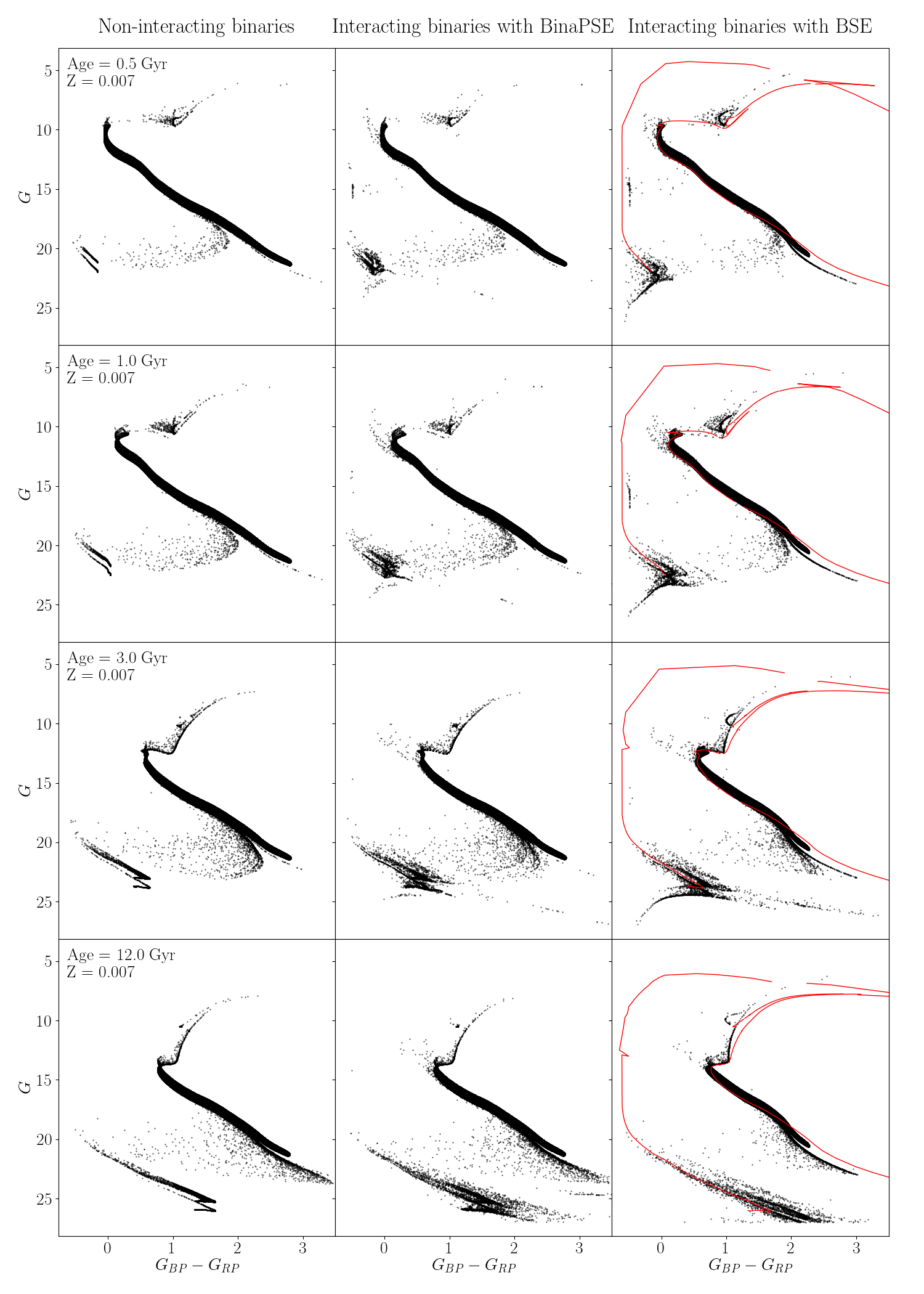}
    \caption{Gaia CMDs of SSPs simulations with initial binary fraction equal to 1.0, metallicity $Z=0.007$, distance 1 kpc, initial mass $10^5\,M_\odot$ and ages 0.5, 1.0, 3.0, and 12.0 Gyr. Left column: simulations with non-interacting binaries. Central column: simulations with BinaPSE. Right column: simulations with BSE. PARSEC isochrones are superimposed to BSE simulations as red solid lines. The isochrones show a discontinuity in the TP-AGB part that is due to the increase of the surface abundance ratio C/O over unity: when C/O becomes greater than 1 then the stellar spectrum of TP-AGB stars changes dramatically \citep[see][for details]{marigo17}.}
    \label{fig:clusters}
\end{figure*}

Simulations of simple stellar populations (SSPs) with initial binary fraction equal to 1.0, metallicity $Z=0.007$, distance 1 kpc, initial mass $10^5\,M_\odot$ and ages 0.5, 1.0, 3.0, and 12.0 Gyr are shown in the CMDs of Fig.~\ref{fig:clusters}. They are build in the Gaia DR2 photometric system. Binary systems are generated by TRILEGAL according to distributions described in Section \ref{sec:probdist}. These SSPs have been simulated in three different ways: as if all binary systems were non-interacting, with the BinaPSE code, and with the BSE code (left, central and right columns of Fig. \ref{fig:clusters}, respectively). For these plots, all binaries are assumed not to be resolved, i.e. the fluxes of the two components in the Gaia filters are summed together, and each point represents either a binary system or single stars resulting from binary interactions.

We first analyse the simulations with non-interacting binaries and compare them with the BinaPSE simulations. In the 0.5 Gyr old SSP we can easily notice the absence of hot-subdwarfs at $G\sim 15$ and of MS stars above the turn-off, when neglecting binary interactions. Moreover, binary interactions lead to an under-populated AGB, especially at younger ages, and to a greater complexity in the WDs region. The latter is due to the presence of He-WDs, which are another peculiarity of binary evolution, and to ONe-WD, which are not included in the simulations without binary interactions. Common features between BinaPSE and non-interacting binaries simulations are: the binary MS, which leads to a double turn-off; some CHeB stars are spread towards bluer colors because of a MS companion; at younger ages CHeB-CHeB binary systems are found at higher luminosity with respect to the other stars or binary systems in the CHeB region of the CMD. Finally, notice the binary CO-WD cooling sequence and the stream of MS-COWD binaries that connects the WDs region with the MS.

On the other hand, Fig.~\ref{fig:clusters} allows us to emphasise the main differences between BinaPSE and BSE synthetic stellar populations. BinaPSE evolution is based on the evolutionary tracks described in Section \ref{sec:binapse_code}, so binary systems lie on or around the corresponding isochrone in a SSP simulation. BSE simulations are based on the evolutionary tracks of \cite{hurley2000} that lead to different isochrones. In order to facilitate comparisons, we superimposed PARSEC isochrones to BSE simulations as red solid lines. The most remarkable differences can be observed in the post-MS phases: the predicted Hayashi line is systematically steeper in BSE than for the PARSEC isochrones. As a consequence, the color distribution in a BSE simulation is shifted towards the blue with respect to BinaPSE simulations. The shape of the CHeB part of the isochrones presents important differences: starting from the 3-Gyr old SSPs, PARSEC isochrones lead to the formation of a compact red clump while BSE simulations always show a blue-loop like shape. 

Low main sequences show different inclinations too. At very low masses the MS of binaries is interrupted before the MS of single stars, because of the lower mass limit of 0.1 $M_\odot$ in both BinaPSE and BSE. For what concerns hot-subdwarfs, we notice that in both kinds of simulations they disappear at older ages, but stellar counts are significantly lower in the BinaPSE case. This fact can be justified with the adoption of different HeMS timescales. The binary systems composed by a HeMS star and a MS star, which are located in the HRD between the HeMS group and the MS, follow the same rule, so in BinaPSE they are present only at very young ages. Finally, the two codes differ also for the predicted stellar counts of ONe-WDs. These WDs lie on a sequence, well populated in BSE simulations with ages up to 3 Gyr, beginning from the tail of CO-WDs cooling sequence and it extending towards the lower left corner of the CMD. ONe-WDs are very few in BinaPSE simulations, highlighting the different IFMR with respect to BSE.

Finally, we find that BinaPSE simulations require at most 3 times the computational time with respect to BSE simulations. However, we consider the accuracy reached by BinaPSE as a good reward for this additional time. Codes have been executed by using a single Intel Core i7-8750H CPU with a clock rate of 2.20 GHz on a personal laptop. We have registered a maximum CPU time of around 10 min for the 12 Gyr old SSP simulated with BinaPSE. 

\section{Testing the HRD-fitting method}
\label{sec:testing}

\begin{table*}
    \caption{The results for 3 exercises of mock CMD fitting (set1 to set3). Mocks are created with the SFR, metallicities, and binary fraction listed in the columns with ``(in)''. The next column then lists the results for the median and 68\% confidence interval, either divided by the input value (in the cases of the SFR and binary fraction), or with the difference between output and input values in the case of metallicities. The second row indicates the total numbers of simulated stars, which is, by construction, comparable with the observed value of 485\,833 stars.}
    \label{tab:mock_solutions_all}
    \centering
    \begin{tabular}{ccccccc}
    \toprule
    & \multicolumn{2}{c}{\textbf{set1}} & \multicolumn{2}{c}{\textbf{set2}} & \multicolumn{2}{c}{\textbf{set3}} \\
    & \multicolumn{2}{c}{\textbf{(535\,995 stars)}} & \multicolumn{2}{c}{\textbf{(461\,418 stars)}} & \multicolumn{2}{c}{\textbf{(422\,307 stars)}} \\
      \cmidrule(lr){1-1} \cmidrule(lr){2-3}  \cmidrule(lr){4-5}  \cmidrule(lr){6-7} 
    age bin & SFR(in) & SFR(out)/SFR(in) & SFR(in) & SFR(out)/SFR(in) & SFR(in) & SFR(out)/SFR(in) \\
    $\logtyr$ & [$10^{-3}\Msun\,\mathrm{yr}^{-1}$] &  & [$10^{-3}\Msun\,\mathrm{yr}^{-1}$] &  & [$10^{-3}\Msun\,\mathrm{yr}^{-1}$] &  \\
      \cmidrule(lr){1-1} \cmidrule(lr){2-3}  \cmidrule(lr){4-5}  \cmidrule(lr){6-7} 
      6.6-7.1 & 0.015 &  1.188 (+0.332/-0.297) & 0.060 &  1.225 (+0.364/-0.323) & 0.135 &  0.801 (+0.148/-0.119) \\
      7.1-7.3 & 0.060 &  1.871 (+0.436/-0.400) & 0.060 &  0.322 (+0.531/-0.241) & 0.120 &  0.494 (+0.523/-0.357) \\
      7.3-7.5 & 0.120 &  1.183 (+0.536/-0.565) & 0.015 &  3.947 (+3.651/-2.863) & 0.075 &  0.516 (+0.744/-0.360) \\
      7.5-7.7 & 0.075 &  0.837 (+0.724/-0.610) & 0.075 &  0.474 (+0.532/-0.359) & 0.045 &  0.614 (+0.816/-0.464) \\
      7.7-7.9 & 0.090 &  0.432 (+0.499/-0.319) & 0.060 &  1.303 (+0.662/-0.744) & 0.015 &  4.223 (+2.886/-2.607) \\
      7.9-8.1 & 0.060 &  1.242 (+0.629/-0.550) & 0.045 &  0.693 (+0.562/-0.409) & 0.060 &  1.599 (+0.682/-0.611) \\
      8.1-8.3 & 0.090 &  0.843 (+0.265/-0.249) & 0.030 &  1.178 (+0.524/-0.593) & 0.120 &  0.764 (+0.185/-0.217) \\
      8.3-8.5 & 0.045 &  1.059 (+0.262/-0.248) & 0.105 &  1.069 (+0.102/-0.104) & 0.030 &  1.431 (+0.367/-0.344) \\
      8.5-8.7 & 0.135 &  1.007 (+0.053/-0.067) & 0.105 &  1.014 (+0.070/-0.079) & 0.090 &  0.878 (+0.076/-0.070) \\
      8.7-8.9 & 0.075 &  1.023 (+0.061/-0.066) & 0.090 &  0.973 (+0.056/-0.050) & 0.030 &  0.954 (+0.114/-0.115) \\
      8.9-9.1 & 0.075 &  1.010 (+0.050/-0.052) & 0.135 &  1.044 (+0.039/-0.037) & 0.120 &  1.067 (+0.028/-0.024) \\
      9.1-9.3 & 0.105 &  0.960 (+0.035/-0.029) & 0.135 &  0.970 (+0.029/-0.033) & 0.075 &  0.992 (+0.030/-0.032) \\
      9.3-9.5 & 0.105 &  0.982 (+0.025/-0.023) & 0.090 &  1.018 (+0.034/-0.035) & 0.135 &  0.983 (+0.019/-0.022) \\
      9.5-9.7 & 0.030 &  0.992 (+0.059/-0.057) & 0.015 &  1.001 (+0.180/-0.170) & 0.135 &  1.012 (+0.027/-0.030) \\
      9.7-9.9 & 0.105 &  1.018 (+0.016/-0.019) & 0.120 &  0.996 (+0.018/-0.019) & 0.120 &  0.992 (+0.023/-0.022) \\
     9.9-10.1 & 0.105 &  1.007 (+0.010/-0.011) & 0.090 &  1.005 (+0.014/-0.012) & 0.030 &  1.000 (+0.030/-0.025) \\
      \cmidrule(lr){1-1} \cmidrule(lr){2-3}  \cmidrule(lr){4-5}  \cmidrule(lr){6-7} 
    & {[Fe/H]}(in) & {[Fe/H](out)-[Fe/H](in)} & {[Fe/H]}(in) & {[Fe/H](out)-[Fe/H](in)} & {[Fe/H]}(in) & {[Fe/H](out)-[Fe/H](in)} \\
    & [dex] & [dex] & [dex] & [dex] & [dex] & [dex] \\
      \cmidrule(lr){1-1} \cmidrule(lr){2-3}  \cmidrule(lr){4-5}  \cmidrule(lr){6-7} 
      6.6-7.1  &  0.2153 &   0.0007 (+0.0035/-0.0036) &  0.4313 & 0.0095 (+0.0037/-0.0037) & -0.0007 &   0.0047 (+0.0035/-0.0038) \\
      9.1-9.3  & -0.0621 &  -0.0016 (+0.0025/-0.0025) &  0.1539 & 0.0024 (+0.0026/-0.0026) &  0.1595 &   0.0018 (+0.0024/-0.0025) \\
      9.9-10.1 & -0.2306 &   0.0018 (+0.0013/-0.0012) & -0.2066 & 0.0002 (+0.0012/-0.0007) & -0.1586 &  -0.0008 (+0.0027/-0.0026) \\
      \cmidrule(lr){1-1} \cmidrule(lr){2-3}  \cmidrule(lr){4-5}  \cmidrule(lr){6-7} 
           & f$_{\text{bin}}$(in) & f$_{\text{bin}}$(out)/f$_{\text{bin}}$(in) & f$_{\text{bin}}$(in) & f$_{\text{bin}}$(out)/f$_{\text{bin}}$(in) & f$_{\text{bin}}$(in) & f$_{\text{bin}}$(out)/f$_{\text{bin}}$(in) \\
      \cmidrule(lr){1-1} \cmidrule(lr){2-3}  \cmidrule(lr){4-5}  \cmidrule(lr){6-7} 
       all ages           & 0.5 & 1.002 (+0.007/-0.007) & 0.9 & 0.998 (+0.004/-0.005) & 0.8 & 1.001 (+0.004/-0.005) \\
    \cmidrule(lr){1-1} \cmidrule(lr){2-3}  \cmidrule(lr){4-5}  \cmidrule(lr){6-7} 
     & & $-\ln\mathcal{L}$ & & $-\ln\mathcal{L}$ & & $-\ln\mathcal{L}$ \\
    \cmidrule(lr){1-1} \cmidrule(lr){2-3}  \cmidrule(lr){4-5}  \cmidrule(lr){6-7} 
    all ages & & 964.5 & & 935.4 & & 986.0 \\
    \bottomrule
    \end{tabular}
\end{table*}


\begin{figure*}
     \centering
     \includegraphics[width=\textwidth]{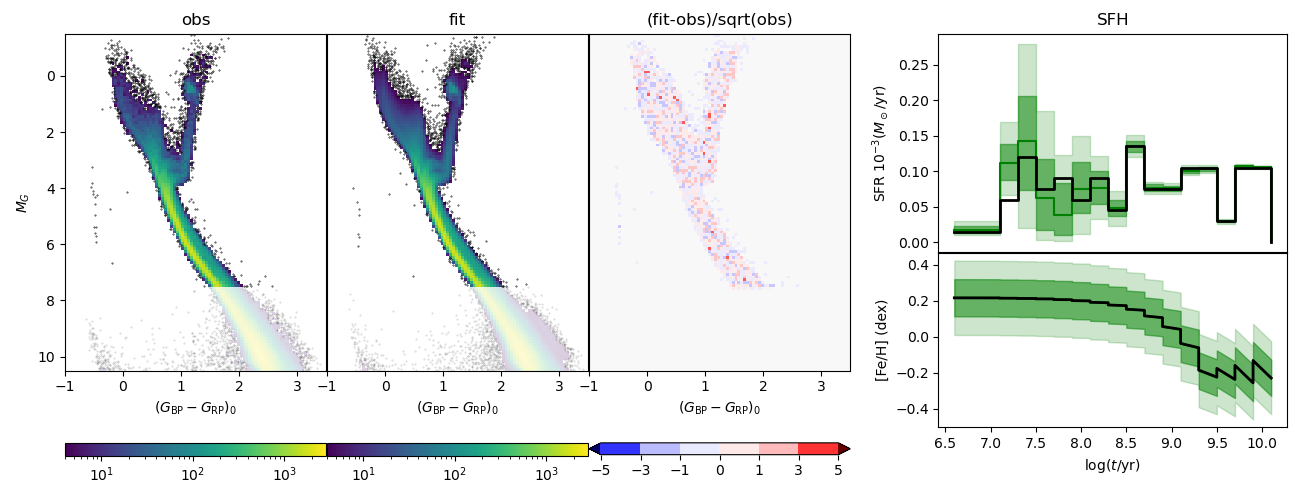} 
     \caption{Results for the test ``set1''. The first two panels show the observed and simulated Hess diagrams, while the third one shows the residuals. The rightmost panels show the SFR$(t)$ (upper panel) and AMR (lower panel) for each age bin with their respective 68\% and 95\% confidence limits, shown as shaded lines. The black line shows the ``true'' solution used to create the observed Hess diagram.}
     \label{fig:summary_set1_noisy}
 \end{figure*}

We describe here a series of experiments aimed at testing the capacity of our method in finding the best-fit solution, and the reliability of the error bars.

The first point to mention is that we have initially tested the code by fitting mock HRDs containing only white noise, at several levels. For instance, we produce a noisy Hess diagram with 100 bins randomly populated from a Poisson distribution with an average rate of 20 stars per bin. This diagram is then fitted with our MCMC code, producing a result of $19.365(-0.427,+0.438)$ for the median and the 68\% confidence interval. The error derived agrees with the $\sigma=0.447$ value expected from a simple analytical formula. For a similar experiment with an average rate of 2 stars per bin, the result is $1.948(-0.134,+0.144)$, whereas the expected error is of $\sigma=0.141$. This kind of test ensures the correctness of the likelihood formula in eq.~\ref{eq:likelihood} -- and hence the correct size of our derived error bars -- even for very low levels of star counts.

Then, we tested the fitting with mock catalogues of the Gaia 200 ~pc sample. These were produced from the same PMs we employed to determine the SFH of the Solar Neighbourhood using known input SFR$(t)$, AMRs and \fbin. At first we tested the code using a flat SFR$(t)$ and AMR, and the code was able to recover the original parameters with very high accuracy. However, this case is not very realistic. To simulate as much as possible a real fitting, we switched to randomly-chosen SFR$(t)$, AMRs, and \fbin\ values. In short, the input models are built with SFR$(t)$ that randomly assume, at any age bin, one of 10 equally-spaced values of SFR between $1.5 \times 10^{-5}\Msun/yr$ and $13.5 \times 10^{-5}\Msun/yr$. These SFR$(t)$ values ensure that the numbers of simulated stars is always similar to the one observed in our Gaia data (485\,833 stars, Sect.~\ref{sec:culling}).
Similarly, the metallicity shifts assume values between $-0.36$~dex and $0.36$~dex with steps of $0.1$~dex, and the \fbin\ assumes any value between $0$ and $1$ with steps of $0.1$.

One of these mock tests, namely ``set1'', is illustrated in Fig.~\ref{fig:summary_set1_noisy}. The first panel to the left shows the ``observed'' Hess diagram, while the second panel shows the model recovered by the code. It is evident that our code determined a very good solution in this case, as highlighted by a very low value of $-\ln \mathcal{L}$ and the very low level of residuals, shown in the third panel. The two rightmost panels illustrate the recovered SFH and their errors. As shown by the top panel, the uncertainties on the SFR$(t)$ from $\logtyr=7.1$ to $\logtyr=8.1$ are significantly larger than for other age intervals. The errors in the output metallicities, however, are always very small, and the same happens for the binary fraction: while the input \fbin\ is 0.5, the recovered value is $0.501\pm0.004$.

Table~\ref{tab:mock_solutions_all} presents all the input values for this model, plus two other similar models, compared to the output values and their confidence intervals. As can be seen, all the output values turn out to agree with input values within their 68\% confidence intervals, with the exception of a couple of SFR$(t)$ values at young ages (for instance, for the age intervals 7.1-7.3 and 7.7-7.9 in ``set1''), which would nevertheless still agree with the input values within their 95\% confidence intervals.  Similar levels of ``mild disagreement'' are also found for the metallicity values at young ages, but in this case the confidence intervals are extremely narrow (of the order of 0.004~dex) -- so narrow that this discrepancy is likely irrelevant compared to those caused by systematic errors in the stellar models. As for the binary fraction, it is always well recovered within the 68\% confidence interval. 

Another point which is apparent in Fig.~\ref{fig:summary_set1_noisy}, is the low degree of correlation in the SFR$(t)$ found for adjacent age bins, especially at old ages. Let us look for instance at the pronounced SFR$(t)$ peak present in the mock data at the age interval $8.5<\logtyr<8.7$, or the drop in SFR$(t)$ at $9.5<\logtyr<9.7$. These marked features did not ``spread'' into neighbouring age bins, in the recovered solution. This is a consequence not only of the ideal conditions at which these mock simulations are performed, but also of the small size of the HRD bins we adopted for the Gaia data: indeed, our 0.2-dex wide age bins produce PMs with HRD sequences typically much wider than the $0.04\,\mathrm{mag}\times0.04\,\mathrm{mag}$ bins in which the HRDs were split. Therefore, there are large numbers of stars in HRD locations (as the MS turn-off regions, or the subgiant branch) that can be reproduced by models in just one age bin. In this case, the correlation in the SFR between adjacent age bins is much weaker than in the case of more distant galaxies, or more distant Gaia samples -- where photometric errors or extinction can create wide PM sequences in the HRD, creating ample superposition between adjacent age bins. 

Overall, these tests show that our code can recover, without significant issues, the parameter set used to build the mock Hess diagrams. It also supports the very small relative error bars found at intermediate and old ages: they are a consequence of the very large star counts present in these mock catalogues at these age intervals, and of the low degree of correlation between adjacent age bins. 

On the other hand, these extremely good results are made possible by the perfect consistency between the stellar models used for build the mock data, and those used in the model fitting. In the real world, this consistency does not exist -- although it is assumed in every single work of CMD fitting in the literature. The larger residuals we find when fitting the real data in Sect.~\ref{sec:method}, are very likely a consequence of systematic errors in the models, that cannot be properly tested with mock data. 

\bsp	
\label{lastpage}
\end{document}